\documentclass[journal]{IEEEtran}
\usepackage{color}
\usepackage{subcaption}
\usepackage{booktabs}
\usepackage{listings}

\usepackage{enumitem}

\usepackage{hyperref}

\ifCLASSINFOpdf
\else
\fi
\usepackage{amsmath}
\usepackage{amssymb}
\usepackage{graphicx}

\usepackage{soul}
\usepackage{makecell}

\begin{document}
%
\title{Digital Twin for Intent-Based Orchestration in 6G}
%
%
%

\author{Marcin~Pakula,
        Lukasz~Kulacz,~\IEEEmembership{Member,~IEEE,}
        Karolina~Lenarska,~\IEEEmembership{Member,~IEEE,}
        Pawel~Kryszkiewicz,~\IEEEmembership{Senior Member,~IEEE,}
        ~Adrian~Kliks,~\IEEEmembership{Senior Member,~IEEE},
        Engin Zeydan,~\IEEEmembership{Senior Member,~IEEE}
\thanks{This work is supported by UNITY-6G project, co-funded
from European Union’s Horizon Europe Smart Networks and
Services Joint Undertaking (SNS JU) research and innovation
programme under the Grant Agreement No 101192650 and
from the Swiss State Secretariat for Education, Research and
Innovation (SERI). The work of Marcin Pakuła was supported in part by the PhD scholarship provided by Siemens Digital Industries Software.}
\thanks{Engin Zeydan is with the Centre Tecnològic de Telecomunicacions de Catalunya, Castelldefels, Spain. Other authors are with Institute of Information and Communication Technologies, Poznan University of Technology, Poznan, Poland. 
Corresponding author: P. Kryszkiewicz, e-mail: pawel.kryszkiewicz@put.poznan.pl}
}

%
%

\markboth{}%
{Shell \MakeLowercase{\textit{et al.}}: Bare Demo of IEEEtran.cls for IEEE Journals}
%



\maketitle

\begin{abstract}
Future sixth-generation (6G) networks will offer a wide range of adaptable, virtualized services across the Radio Access Network (RAN), Core Network (CN), Transport Network (TN), and emerging domains such as Artificial Intelligence (AI). Meanwhile, stakeholders will express highly heterogeneous intents describing the desired end-to-end (E2E) communication services. Mapping these intents to concrete service configurations is challenging because of the large number of possible service realizations, dynamic resource availability, and ambiguous intent interpretation. Existing orchestration frameworks lack mechanisms to reliably evaluate alternative service allocations before deployment, which is critical given the high cost and long instantiation times of virtualized network services. To address these challenges, this paper introduces a Digital Twin-assisted orchestration framework integrated into a unified, service-based 6G architecture. Each service provider exposes a DT that models its internal behavior while preserving privacy, enabling the Inter-Domain Management and Orchestration (IDMO) entity to assess the performance and feasibility of candidate intent-service mappings without impacting the live network. We develop a probabilistic model of DT outputs, an online estimator of systematic DT prediction errors, a signaling workflow for intent processing, and an optimization problem to select the most suitable service configuration under uncertainty.
By evaluating the predicted resource demands and interactions among intents sharing a service, the proposed framework can reduce conservative resource overprovisioning and enable more resource-efficient intent-to-service mapping. 
System-level simulations, including a disaster recovery scenario, demonstrate that DT-assisted orchestration enables resource-efficient intent-to-service mapping, enhancing end-to-end service continuity in complex 6G environments.
\end{abstract}


%
\IEEEpeerreviewmaketitle

\section{Introduction}
\IEEEPARstart{T}{he} transition to sixth-generation (6G) networks introduces unprecedented requirements for scalable, open, and sustainable service provisioning across heterogeneous domains, including the Radio Access Network (RAN), Core Network (CN), and Transport Network (TN). Emerging applications, such as extended reality, disaster response, and industrial automation, require reliable low-latency communication and intelligent service provisioning over distributed infrastructures. Since the introduction of 5G, communication networks have increasingly adopted software-defined networking (SDN) and virtualized network functions (VNFs) \cite{Kreutz2015, Mijumbi2016}. By decoupling network functions from the underlying infrastructure, these technologies enable high-level network services (NSs) to be deployed without detailed knowledge of the available hardware and connectivity. Mapping service requirements to physical, computational, and connectivity resources is performed by management and orchestration entities, such as ETSI NFV MANO \cite{ETSI_GR_NFV-MAN_001_v1.2.1}. This service-oriented approach also laid the foundation for applying Service-Based Architecture (SBA) to telecommunication networks \cite{Chatras2018}, where architectural components expose services rather than fixed reference points. 

Several management and orchestration solutions have been developed for individual domains, such as the RAN and CN. They either extend ETSI NFV MANO \cite{Makris2018, Breitgand2021, Makaya2015} or adopt broader approaches, including Open Source MANO \cite{etsi_osm_docs} and ETSI Zero-touch Service Management (ZSM) \cite{etsi_zsm_002}. However, existing frameworks still provide limited support for dynamic cross-domain orchestration and end-to-end (E2E) service continuity \cite{ETSI_GR_NFV-MAN_001_v1.2.1, wang2020design}. This limitation is particularly relevant to Intent-Based Networking (IBN) \cite{Leivadeas2023}, where a stakeholder expresses a service request, referred to as an intent, in natural language. The intent is subsequently translated, e.g., using a Large Language Model (LLM), into technical requirements that must be mapped to services available across multiple domains. 

Recent work in \cite{blanco2025unity,unity6g_d23_2026} introduces a unified 6G architecture integrating Open RAN \cite{Polese2023,Hoffmann2024}, distributed intelligence \cite{Abdulrahman2021}, and AI-native orchestration \cite{Wu2022,Hoydis2021}. It comprises functional domains providing RAN, CN, non-terrestrial network (NTN), non-public network (NPN), artificial intelligence (AI), and Digital Twin (DT) services. These domains are coordinated by a centralized Inter-Domain Management and Orchestration (IDMO) entity. The IDMO translates stakeholders' intents into E2E requirements and maps them to the services exposed by the available domains.

Such intent-to-service mapping is challenging because it requires accurate and up-to-date information about distributed network resources. A single E2E intent may involve several alternative implementations combining RAN, CN, TN, or NTN services. Although this problem can initially be viewed as an assignment or knapsack problem, neither the available resources nor the resources required by a service are precisely known in real networks. While \cite{Feng_TNSM_2026} addresses a related intent-driven resource-allocation problem by estimating the available resources (i.e., the knapsack size) using machine-learned prediction, it does not explicitly model, e.g.,interactions among co-allocated intents.

Additional uncertainty results from ambiguities in translating natural-language intents into technical requirements. 
Moreover, several allocations may satisfy an intent but differ in resource consumption and deployment cost. Deploying a virtualized E2E service can be time-consuming and resource-intensive \cite{unity6g_d23_2026}, so we cannot efficiently test multiple candidate allocations in the physical network. Consequently, the mapping decision should account for the expected performance of candidate services before their deployment. 

A Digital Twin (DT) is a digital representation of a physical system, in this case, a communication service provided by a wireless network \cite{2021DTSurvey}. DTs provide a risk-free environment in which candidate configurations can be evaluated without affecting the physical network, e.g., to prevent conflicts in configurations \cite{Polverini_TNSM_2026}. The general idea of using DTs for intent-based networking was introduced in \cite{wang2025revolutionizing}, with an emphasis on LLM-based intent translation. However, that work assumes a perfect DT and does not formulate the intent-to-service assignment problem. In \cite{de2024optimizing}, DT-assisted orchestration is formulated as a many-to-many matching game between service requests and network domains, optimizing Key Performance Indicators (KPIs) and Key Value Indicators (KVIs). However, the proposed approach neither considers ambiguities in intent-to-KPI translation nor accounts for DT prediction errors, while the employed DT models rely on simplified analytical expressions. A more realistic hierarchical DT architecture is proposed in \cite{jia2025hierarchical}, where AI models trained on network data first identify problematic network regions and then support detailed local optimization. Nevertheless, that work focuses on user association and power allocation rather than intent-to-service assignment. 

In this work, we propose using DTs to evaluate prospective intent-to-service assignments before their deployment in the physical network. Each service is assumed to expose its DT functionality to the IDMO through an Application Programming Interface (API). Since the service provider has the most complete knowledge of its service architecture, configuration, and resource-allocation mechanisms, its DT can provide the most reliable estimate of the performance achievable for a given intent-to-service assignment. As an additional advantage, the provider may adapt both the service configuration and its DT model to the requirements of the evaluated intent. Only high-level KPIs are exposed to the IDMO, thereby preserving the confidentiality of the provider's internal configuration. Depending on service interdependencies, the IDMO may use independent service-specific DTs or a unified DT framework to jointly evaluate interacting services \cite{Hindi_OJCOM_6gTWIN_2026}. In either case, the internal DT implementation remains hidden to the IDMO. 
Moreover, by jointly evaluating intents sharing the same service, the proposed approach can reduce the conservative resource overprovisioning required by conventional assignment methods and thereby improve resource utilization.
A major challenge is the inevitable mismatch between DT predictions and the performance observed after service deployment. We explicitly incorporate this mismatch into the IDMO decision-making process in two ways. First, DT outputs are represented probabilistically to account for uncertainties arising from incomplete intent specifications, unknown propagation conditions, and imperfect DT-network synchronization. Second, systematic prediction errors are estimated separately for each service and KPI. The resulting correction allows the orchestrator to penalize persistently optimistic DT predictions, including cases in which a provider unintentionally or deliberately overestimates its service performance.

Compared with the existing literature, the main contributions of this paper are as follows:
\begin{itemize}[leftmargin=*] 
\item We propose an imperfect-DT-assisted intent-management framework for service-based architectures, including its functional components, message exchange, and allocation procedure. 
\item We introduce a probabilistic intent-fulfillment model that accounts for uncertainties in intent-to-service translation, and systematic DT prediction errors. 
\item We develop an online estimator of systematic service- and KPI-specific DT prediction errors. The estimator allows the IDMO to penalize systematically optimistic predictions while averaging out random differences between predicted and observed KPI fulfillment.
\item We formulate intent-to-service allocation as an optimization problem that maximizes the minimum KPI-fulfillment probability across all intents and KPI requirements. 
\item We evaluate the framework under different network conditions, including infrastructure failures, and demonstrate how DT predictions and their online correction improve the reliability of intent fulfillment. 
\end{itemize}

The remainder of the paper is organized as follows. Section~\ref{sec:sba} presents the service-based architecture, E2E service-creation workflow, and DT-assisted evaluation process. Section~\ref{sec:service-allocation} formulates the allocation problem and its solution. Section~\ref{sec:scenario_results} describes the evaluation scenarios and discusses the results. Finally, Section~\ref{sec:conclusions} concludes the paper.


\section{Digital Twin-aided Service-Based Architecture}
\label{sec:sba}
For many years, the concepts of network virtualization and software-defined networking have attracted considerable attention. The CN part of 5G networks has been standardized in a way that promotes a service-based approach, in which most network functions - such as the Access and Mobility Management Function (AMF), Session Management Function (SMF), Authentication Server Function (AUSF), and Network Exposure Function (NEF), among others - are represented as functions connected by a dedicated bus \cite{3gpp_ts_23501,3gpp_ts_29500_rel17}. In essence, the key control-plane functions are defined as part of the SBA. However, with the advent of 6G solutions, discussions about the new network architecture increasingly focus on incorporating openness, distributed intelligence, and resource orchestration. Several Standards Development Organizations promote key elements of unified future networks \cite{unity6g_d23_2026}. ETSI focuses on NFV, MANO, and AI-assisted automation \cite{ETSI_GR_NFV-MAN_001_v1.2.1,chafika2021distributed}. TM Forum develops cloud-native architectures and intent models \cite{tmforum_tr299_intent_specification,tmforum_tr290_intent_common_model}. O-RAN specifies an open, disaggregated RAN \cite{Polese2023}. Finally, ITU-T develops standards for autonomous networks, cross-domain orchestration, and AI integration \cite{itu_y3172_2019}.

\begin{figure}[htp!]
    \centering
    \includegraphics[width=0.8\columnwidth]{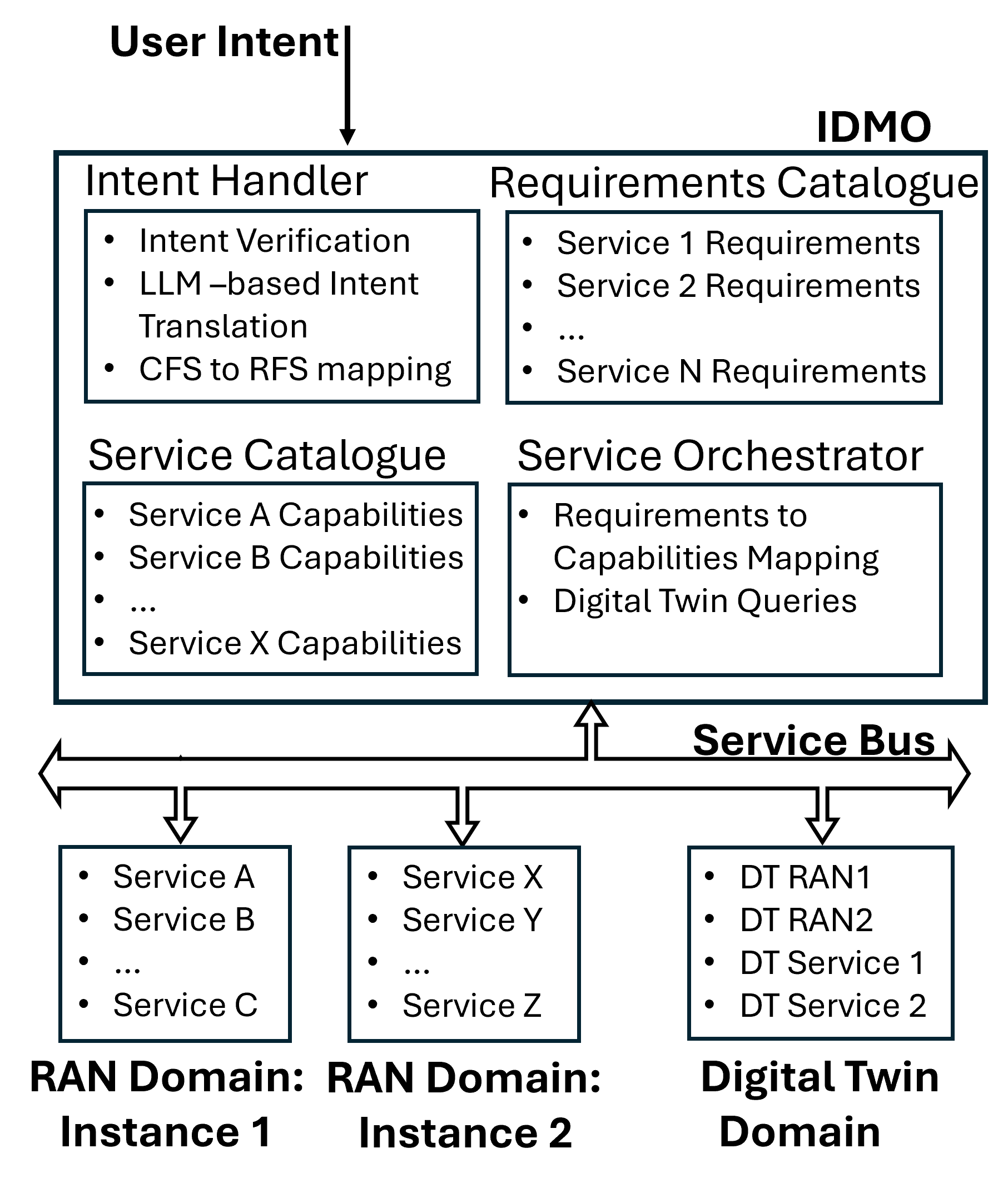}
    \caption{Multi-domain Service-Based System Architecture.}
    \label{fig:architecture}
    \vspace{-.5cm}
\end{figure}

 
These aspects are expected to be reflected in the final 6G architecture. Building on this observation, our research examines a fully virtualized SBA that incorporates an intent-based approach, as shown in Fig.~\ref{fig:architecture}. From a high-level perspective, the architecture consists of three key parts: (1) a set of domain instantiations, (2) the Inter-Domain Manager and Orchestrator (IDMO), and (3) the Service Bus connecting them.

In our approach, a domain is the logically separated part of the overall communication system that provides specific functionalities to the system. For example, the radio access domain (RAN) will be responsible for delivering functionalities and services related to radio communication, whereas the NTN domain will cover functionalities related to satellite and other non-terrestrial communication. Every \textit{domain}, by definition, has its own functions, resources, and control mechanisms, and is typically managed independently of other domains, and exposes interfaces to the outside world. As the domain is primarily a theoretical concept, not tied to a particular implementation, in the case of a RAN provided by a specific mobile network operator (MNO), it is more appropriate to refer to a \emph{domain instantiation}. This means that we assume the possible coexistence of several domain instantiations, as shown in Fig.~\ref{fig:architecture}. Through the service bus, each domain exposes the list of services that can be offered as part of the E2E services to the client. Here, a \textit{service} is understood as a set of domain-specific functionalities delivered to a user, realized by orchestrating resources and functions within a single domain, while an \textit{E2E service} refers to a service that spans multiple domains and is composed of a set of domain services. In the literature, the E2E service is sometimes called an integrated service or a business-level service, as it may be a concatenation of services originating from different domains. Hereafter, we use the term E2E service to emphasize its cross-domain nature.  Each service provider may be interested in delivering various services to the end client to maximize its gains and revenues. For example, the MNO, as the RAN service provider, may offer localization services, mobile edge computing services, or various communication services (described, for example, by different throughput and latency metrics). Clearly, each service will have specific deployment constraints and costs. Thus, to make the service available to the end client, the service provider must expose the list of available services to the IDMO. As shown in Fig.~\ref{fig:architecture}, the domains deliver the service descriptions to IDMO, which ingests into its \textit{service catalog} (or service pool). Various forms of service representation may be used, such as an XML document or a JSON catalog, but this aspect is outside the scope of this paper. 

The interested stakeholder (e.g., a virtual MNO) provides the service request as an \textit{intent} expressed in natural language \cite{etsi_ts_128312}. The IDMO entity receives and translates the incoming intent within the Intent Handler through a two-stage process, following TM Forum guidelines \cite{tmforum_sid_gb922} and facilitated by a LLM. First, the intent is mapped to a Customer Facing Service (CFS), which defines the service in terms of user-visible characteristics and high-level requirements. Subsequently, the CFS is translated into a Resource-Facing Service (RFS) for the CN and RAN domains, specifying the technical parameters and resource-level configurations required for network realization. These requirements, represented in a structured format such as a JSON file, are then stored in a \textit{Requirement Catalogue} (analogous to how domain services are maintained in the Service Catalog). To deploy and run an E2E service for the User, the Service Orchestrator maps the requirements to the services. Following the SBA approach, communication between the IDMO and the domain instances is carried out through the dedicated bus \cite{unity6g_d23_2026}, whose definition is outside the scope of this paper. Once the Service Orchestrator determines the mapping between service requirements and capabilities, it instantiates these services in the domains. Each domain is then responsible for realizing (deploying, launching) the new service. Very often, the service within the domain is deployed as a network slice. At the same time, the client is informed about the success or failure of the end-to-end service deployment. To ensure the quality of deployed services, IDMO continuously monitors the selected KPIs reported by the domain instances. If a significant degradation in service quality is detected or reported, IDMO initiates the procedure to reallocate the service associated with the relevant intent.

However, the optimal process for associating intents with services is complex. It can be treated as a generalized form of the well-known knapsack problem, where the knapsack size (available network resources) varies over time, as does the size of each element (resources needed by a given intent), due to the nature of wireless networks. Here, the number of parameters to consider and their potential interdependencies make the problem extremely complex. Thus, in this work, we propose using the DT to evaluate the potential associations. As illustrated in Fig.~\ref{fig:architecture}, the DT constitutes a separate domain related to digital-twin-specific functionalities. These services are then available to IDMO, which uses them by requesting the DT instance to run a dedicated DT service to evaluate the performance of the prospective association between the requirements and capabilities. Despite advancements in specific domains, a comprehensive E2E service management framework remains largely undefined in current research. Similarly, the intent-based management approach proposed in \cite{Mekrache2024} primarily addresses reliable intent interpretation using LLMs but does not provide a complete E2E architectural framework.



Current frameworks (e.g., \cite{ETSI_GR_NFV-MAN_001_v1.2.1, benzaid2020ai, wang2020design}) typically operate in isolation, target individual domains (e.g., RAN or CN), and rely on static interfaces and hierarchical control structures that limit scalability and flexibility. This approach hinders seamless end-to-end orchestration and dynamic service continuity, especially when resources must be reallocated across domains. Moreover, these frameworks lack standardized mechanisms for integrating AI-driven closed-loop control across domains, which leads to the underutilization of distributed intelligence. These limitations motivate the development of a Service-Based Management Architecture (SBMA), which enables intelligent, cross-domain, sustainability-aware orchestration, as presented in \cite{blanco2025unity}. 








\subsection{ Intent mapping to service KPIs}

Mapping natural-language intents to service KPI requirements is a key function of autonomous networks. It enables the transformation of user goals, represented by natural-language intent, into measurable technical parameters required for proper infrastructure configuration. The emergence of LLMs offers promising capabilities for bridging this gap, as their natural language understanding and generation abilities allow for smooth translation of high-level human requirements into precise technical policies. 

LLM-based intent translation in 5G/6G networks mainly follows two paradigms: prompt engineering and parametric fine-tuning. Prompt engineering leverages techniques such as multi-stage mapping (e.g., translating business intents first to Customer Facing Services - CFS and then to Resource Facing Services - RFS), Chain-of-Thought (CoT) prompting, and hierarchical multi-agent architectures to drive intent translation without altering underlying model weights \cite{dinhtowards2025, fonseca2024intent, jiang2026agentic}. On the other hand, parametric fine-tuning techniques, including Low-Rank Adaptation (LoRA), Quantized LoRA (QLoRA), and Proximal Policy Optimization (PPO), often combined with Retrieval-Augmented Generation (RAG), are used to achieve deeper semantic alignment with communication standards and eliminate hallucinations \cite{wang2026lamio, habib2025generative}.

In addition, LLMs can be used to make domain selection and service placement decisions \cite{jiang2026agentic}. However, relying purely on LLMs for end-to-end orchestration introduces notable drawbacks, such as non-deterministic execution, high computational overhead, potential latency bottlenecks, and a lack of real-time network state awareness. To mitigate these limitations, our approach restricts the role of the LLM strictly to intent translation: mapping high-level intents to specific KPI requirements, while leaving the domain matching, service mapping, and resource allocation to be dynamically evaluated using the Digital Twin.

\subsection{Digital Twin for Intent-Service evaluation}

The main objective of incorporating a DT within the network orchestrator is to enable accurate prediction of service performance while supporting a given intent. To achieve this, each service should ideally be associated with a dedicated DT instance. If such a dedicated DT is not available, the IDMO may rely on a simplified or generic service model. However, this approximation may introduce larger modeling errors, making such services less favorable during the selection process. Given a complete set of DTs representing all candidate services, the IDMO can perform DT-based evaluations of different intent-service mappings. Since these mappings are mutually independent, they can be simulated in parallel, which significantly reduces the overall computation time. 

Independent services can be evaluated using separate DTs, whereas interacting services require joint evaluation to capture effects such as shared-resource interference. This may be handled internally by a service provider or through a unified DT framework \cite{Hindi_OJCOM_6gTWIN_2026}, with the IDMO accessing only the resulting KPI distributions through a standardized API.
To ensure optimal operation, KPIs associated with each intent must be continuously monitored and evaluated within the IDMO to enable informed decision-making and adaptive orchestration.

\section{Service allocation in IDMO}
\label{sec:service-allocation}
In addition to intent translation, one of the key responsibilities of the IDMO is the effective utilization of available resources offered by the domains while maximizing intent owners' satisfaction. It can be achieved through precise and intelligent mapping of intent requirements to domain capabilities. Moreover, IDMO should adjust the current intent-service allocation whenever the observed quality of the deployed services falls below the assumed threshold. The most obvious cases when the assignment process runs are:
\begin{itemize}
    \item When a new or modified intent is received (one can call it a top-trigger, as it originates from the client);
    \item When the service characteristics change, e.g., at least one intent does not maintain the Service Level Agreement (SLA)/ QoS requirements (analogously, one can call it bottom-trigger, as the need for mapping modification originates from the underlying domains). Observe that this can result from either a low-level intent change (e.g., users moving within a network) or a service provider reconfiguring its network, such as adjusting its network parameters to support other services or incoming intents. 
\end{itemize}

\begin{figure}[htp!]
        \centering
        \includegraphics[width=0.8\columnwidth]{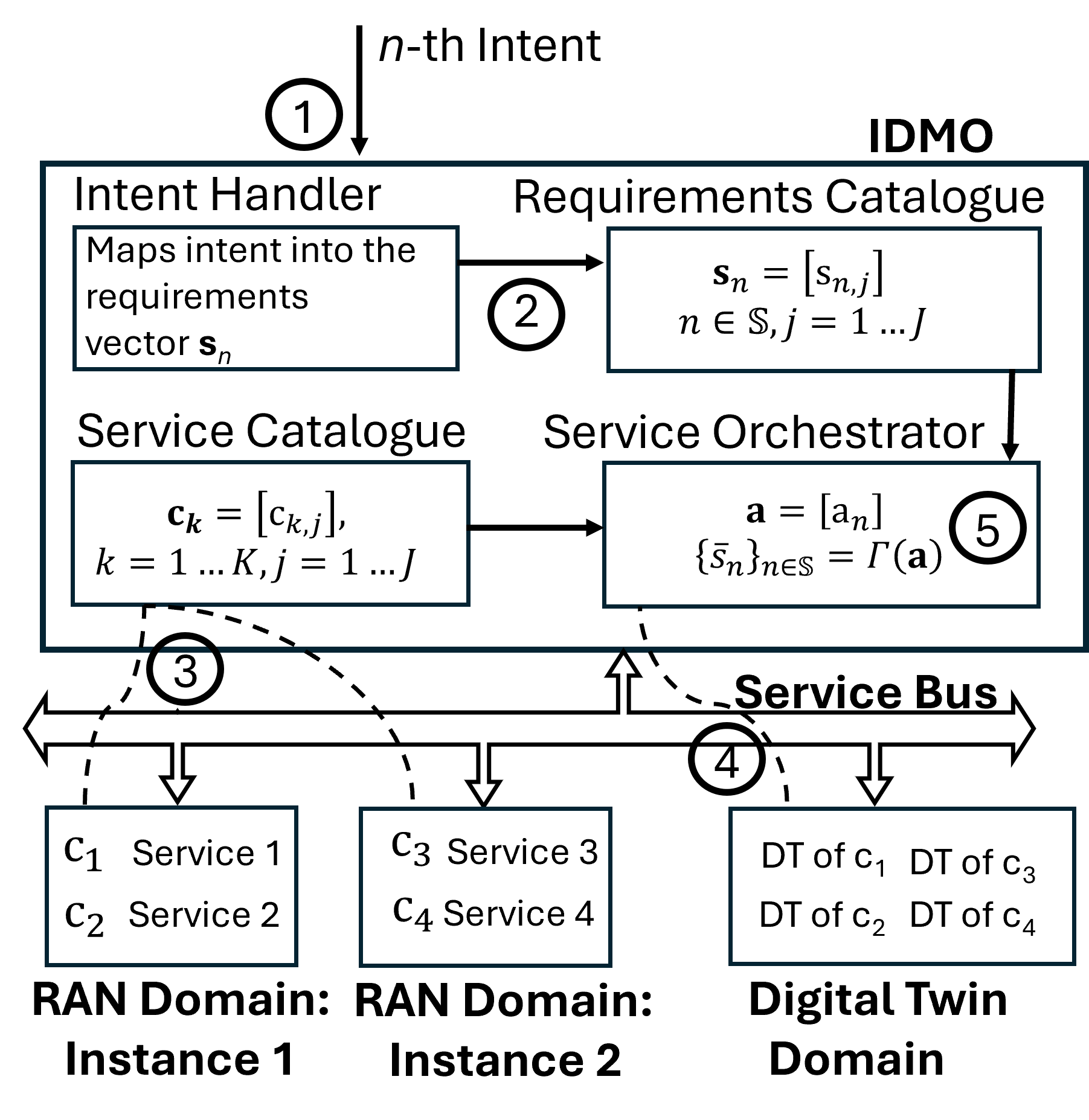}
        \caption{ Message flow between system items during the intent-service allocation process. Mathematical notations explained in Sec. \ref{sec_system_model}.}
        \label{fig:messageflow}
    \end{figure}

The proposed message flow between the specific entities during the intent-service allocation process is illustrated in Fig. \ref{fig:messageflow}. First, the user provides the intent to the IDMO (1). Next, the IDMO processes the intent by translating it and storing its representation as a set of service requirements in the catalog (2). In parallel, all service providers expose their services by communicating their service capabilities to the IDMO through the service bus (3). Moreover, each service provides its digital representation in the form of a Digital Twin to the DT domain (4). The Service Orchestrator tests various intent-service assignments using the DT, first rejecting impossible options based on capability vectors stored in the Service Catalogue (5). The best allocation (the proposed objective function is discussed in the next subsection) is then deployed in the required domain instances. Afterward, the actual network statistics can be observed.

\subsection{System model}
\label{sec_system_model}

Let $\mathbb{S}$ denote the set of identifiers for intents that are to be mapped and allocated to specific network services. Therefore, the total number of active intents to be allocated is $|\mathbb{S}|$. For the $n$-th intent, with $n \in \mathbb{S}$, a multidimensional variable $\mathbf{s}_n = [s_{n,j}]$ defines its requirements (e.g., the number of UEs to be served, the required throughput, or the maximum latency, each in a separate variable dimension), where $j$ denotes the requirement index. There are $K$ available services in total. The $k$-th service is described by a capabilities vector $\mathbf{c}_k$ with a structure identical to the intent requirements $\mathbf{s}_n$. This vector specifies maximum service performance, for example, the maximum UE throughput achievable under perfect propagation conditions, and can be used to filter out infeasible allocations without testing them using the DT. This can save some optimization time and energy required to run the DT. While more than one intent can be assigned to a single service (e.g., more than one group of users can be served by the same MNO resources), there is a maximum of $K^{|\mathbb{S}|}$ possible allocation strategies. The optimization variable is $a_n\in \{ 1,...,K\}$ equal to the index of a service selected for the $n$-th intent. The allocation for all considered intents is denoted by  $\mathbf{a} = \{ a_n \}_{n \in \mathbb{S}}$. The DT is used to answer the question on achievable KPIs when running a given intent-service allocation. In other words, the DT can be treated as a function denoted by $\Gamma$ that takes as input the proposed intent-service allocations $\mathbf{a}$ and outputs the achievable KPIs for the $n$-th intent, $\mathbf{\bar{s}}_n$, with the same structure as $\mathbf{{s}} _n$, resulting in:
\begin{equation}
    \{\mathbf{\bar{s}}_n\}_{n\in\mathbb{S}}=\Gamma(\mathbf{a}).
\end{equation}
Function $\Gamma(\mathbf{a})$ represents the most general case in which the achievable KPIs may depend on the complete allocation vector, for example, because several intents or services share resources or generate mutual interference. This formulation does not require the IDMO to construct or internally execute a monolithic DT. If the services use independent resources, $\Gamma()$ can be factorized into independent service-specific functions, e.g., $\Gamma_k(\mathbf{a})$ for service $k$. A function $\Gamma_k()$ needs to be invoked only if at least one intent is assigned to service $k$. Conversely, interactions among tightly coupled services may be modeled jointly by a unified DT framework\cite{Hindi_OJCOM_6gTWIN_2026} or internally by their common provider. From the IDMO perspective, all implementations expose the same API: the proposed allocation is provided as input and the corresponding KPI distributions are returned as output, and any internal DT structure, e.g., using advanced network simulators or simple look-up tables, is acceptable. 


Most importantly, after selecting one intent-service allocation, the actually achieved KPIs are verified in a real network.
 Let $s^{\mathrm{obs}}_{n,j}$ denote the value of the $j$-th KPI observed in the physical network for intent $n$. Its observed fulfillment is represented by a binary variable 
\begin{equation} 
y_{n,j} = \begin{cases} 
1, & \text{if } s^{\mathrm{obs}}_{n,j} \geq s_{n,j},\\ 
0, & \text{if } s^{\mathrm{obs}}_{n,j} < s_{n,j}. 
\end{cases}
\label{eq_fullfilment_flag}
\end{equation} 
where $y_{n,j}=1$ indicates that the KPI requirement is satisfied, whereas $y_{n,j}=0$ indicates its violation. A violation means that either the DT estimate is incorrect, allowing the IDMO to rerun the assignment and penalizing the DT of the service for being too optimistic (which will be covered in the next subsections), or that there are not enough resources in the network to serve all intents. In the other case, IDMO should inform the “worst-case” intent provider and either reject it or renegotiate its requirements. 


\subsection{DT output modeling}
\label{sec_DT_output}

Most importantly, the DT has to estimate each intent KPI without full knowledge. On the one hand, the intent definition may not be specific, e.g., we know there will be 10 UEs, but their exact locations are unknown. We call this \emph{intent-originating randomness}. 
Additionally, the environment itself can introduce some randomness, e.g., even for UEs at known locations, the propagation channel can change as a result of other objects’ movement. We call it \emph{environment-originated randomness}. Because of this randomness, $\mathbf{\bar{s}}_n$ should be treated as random variables and described by their cumulative distribution functions. This is in line with the proposal in \cite{ruah2023bayesian}. 

The problem is how the DT can assess its KPI estimation distribution when a given intent has never been served by a given service. For this purpose, \emph{randomization}, a tool well known in robotics for simulation-to-reality mapping \cite{tobin2017domain}, can be used. This will require multiple random DT instances to run in parallel to generate a KPI distribution. Moreover, it should be noted that KPIs in a mobile network can vary on a micro time scale, for example, due to the scheduling of different UEs in different time slots. Therefore, real network observations should be conducted over a longer period, such as a few minutes, to estimate actual KPI variability. 
Thus, the distribution should be estimated both across multiple DT realizations and within a limited time window. Three DT randomization states are expected:
\begin{itemize}[leftmargin=*]
    \item \textit{"Random" Case - A given intent-service pair is never used:} This state results from a combination of \emph{intent-originating randomness} and \emph{environment-originated randomness}. Thus, the highest randomness is expected. The DT must consider the entire range of random variables, such as those denoting users' locations.
    \item \textit{"Exact" Case - A given intent-service pair is currently used:} In this case, the lowest KPI variance is expected, as the DT should be updated accurately for a given network state. Nevertheless, temporal variability remains because the DT cannot perfectly predict future UE behavior, even over the short interval between consecutive synchronizations with the physical network. Therefore, the DT KPIs should be modeled as random variables using observations collected over a preceding time window. Moreover, although the intent-service pair under consideration is currently in use, other intents included in the tested DT allocation may be new, introducing additional variance to the KPIs of the \emph{old} allocated intent-service pairs.
    \item \textit{"Partial knowledge" Case - A given intent-service pair was previously used:} 
    Immediately after a given intent-service pair allocation is terminated, the DT can be expected to have full state information (subject to the limitations described in the "Exact" Case). For example, DT knows the path loss of UEs, their terminal capabilities, and so on. Therefore, the variance of KPI estimation should be low. However, as time passes, the DT cannot accurately update all of its state information; for example, the path loss of each UE becomes unknown as they move through space. Consequently, the variance should increase, although its maximum value should be lower than in the “Random” Case, as most of the \emph{intent-originating randomness} is not present.
\end{itemize}

\noindent
 In addition to randomness originating from the environment and from intent, the DT output may contain a systematic prediction error caused by, for example, inaccurate DT models, delayed synchronization, or intentionally biased DT behavior, such as increasing the probability of selecting a given service to increase the operator's revenue. This error can be estimated only after a new allocation is deployed and the intent fulfillment from (\ref{eq_fullfilment_flag}) is calculated. The resulting service-specific error estimates can then be used by the Orchestrator to systematically penalize optimistic DT predictions in subsequent allocation decisions.


 \subsection{Problem formulation} \label{sec_problem_formulation}

The IDMO has to find services that support the $n$-th intent $a_{n}$. While the DT outputs random variables ${\bar{s}}_{n,j}$ the aim should be some probabilistic goal. One option is to maximize the probability that all the KPIs of all intents are met, i.e., 
\begin{align}
   \max_{a_{n}:n\in\mathbb{S}} & \Pr\!\left(\prod_{n,j} \left({\bar{s}}_{n,j} \geq s_{n,j}\right)\right). 
\end{align}
Here, it is assumed that the requirements are lower-bounded, i.e., the achieved KPI must be greater than or equal to the requirement. For some KPIs, such as delay, the reverse inequality denotes successful allocation. For notational simplicity, we will not include this in the optimization problem. This can be achieved by changing the signs of both sides of the inequality; for example, the delay and the delay constraint will become negative numbers.  The above would require knowing the multidimensional distribution or at least the conditional probabilities, i.e., 
\begin{align}
\Pr\!\!\left(\!\!\prod_{n,j} \!\left({\bar{s}}_{n,j} \!\geq\! s_{n,j}\right)\!\!\right)\!\!\!=\!\! \Pr\!\left( {\bar{s}}_{1,j}\! \!\geq\!\! s_{1,j}\!\right)\!\Pr\!\left( {\bar{s}}_{2,j} \!\!\geq\!\! s_{2,j}\!\!\mid\!\! {\bar{s}}_{1,j} \!\geq\!\! s_{1,j}\right)... 
\end{align}

When all KPI-fulfillment events are independent, e.g., do not cause interference to each other, it can be simplified to:
\begin{align}
   \max_{a_{n}:n\in\mathbb{S}} &\prod_{n,j} \Pr( {\bar{s}}_{n,j} \geq s_{n,j})
\end{align}

However, although each intent is concerned only with meeting its own requirements, it is reasonable to maximize the probability that the "worst"/ most demanding intent is served correctly. This is equivalent to maximizing the minimum success probability across all intents and all KPIs, i.e.,
\begin{equation} 
\max_{a_n:n\in\mathbb{S}} \min_{n,j} \Pr\!\left(\bar{s}_{n,j}\geq s_{n,j}\right). 
\label{eq:max_min_goal} 
\end{equation}

 The objective in \eqref{eq:max_min_goal} assumes that the probabilities provided by the DT are correctly calibrated. However, in addition to randomness originating from the environment and from intent, the DT output may contain systematic errors caused, for example, by intentionally optimistic service-specific DT.
Such errors can be evaluated after an allocation is deployed by comparing the predicted fulfillment probabilities with physical network observations. Most importantly, the estimator must be able to quantify the systematic error while filtering out randomness originating from the environment and from intent.

Let $\widehat d_{k,j}^{(\ell)}$ denote the estimated probability prediction error of service $k$ for KPI $j$ after its $\ell$-th observation. Such an observation becomes available only after an intent has been assigned to the service and the resulting performance has been measured in the physical network. When service $k=a_n$ is deployed for intent $n$, the estimate is updated as 
\begin{equation} 
\widehat d_{k,j}^{(\ell)} = (1-\alpha)\widehat d_{k,j}^{(\ell-1)} + \alpha \left[ \Pr\left(\bar{s}_{n,j}\geq s_{n,j}\right)-y_{n,j} \right], \quad k=a_n, \label{eq:dt_prediction_error} 
\end{equation} 
where $0<\alpha\leq1$. The instantaneous prediction error $\Pr\left(\bar{s}_{n,j}\geq s_{n,j}\right)-y_{n,j}$ belongs to $[-1,1]$, and a persistently positive estimate $\widehat d_{k,j}^{(\ell)}$ indicates that the DT of service $k$ systematically overestimates the probability of satisfying KPI $j$. For a correctly calibrated DT, $\mathbb{E}[y_{n,j}]$ equals the predicted fulfillment probability $\Pr\left(\bar{s}_{n,j}\geq s_{n,j}\right)$. Consequently, random variations originating from the environment and individual intents are gradually averaged out, whereas a persistent service-specific prediction error remains reflected in $\widehat d_{k,j}^{(\ell)}$. 
The estimate is maintained per service and KPI rather than per intent, assuming that prediction errors are sufficiently similar across intents evaluated using the same service DT. This aggregation represents a trade-off between context specificity and convergence speed. Because observations are available only after service deployment, context-specific estimates would receive fewer updates and might not converge reliably. Aggregating observations enables faster detection and penalization of systematically optimistic DT predictions, while variations in operating conditions help average out random errors. Initially, $\widehat{d}_{k,j}^{(0)}=0$, so the correction does not affect allocation before any observations are collected. The estimate is updated only when service $k$ is deployed and does not decay over time. Although a forgetting mechanism could accommodate subsequent DT improvements, it could also cause a systematically inaccurate service to be repeatedly selected after its penalty decreases. Conversely, retaining the estimate may continue to penalize a DT that has been improved. The appropriate approach therefore depends on the deployment. 
While the main purpose of estimating $\widehat d_{k,j}^{(\ell)}$ is to penalize bad quality or intentionally optimistic DTs, we can use the positive-part operator $[x]^+=\max(x,0)$. 
Accounting for systematic DT prediction errors, the intent-service allocation problem is formulated as \begin{subequations} 
\label{eq_opt_problem} 
\begin{align} 
\max_{a_n:n\in\mathbb{S}} \quad & \min_{n,j} \left[ \Pr\!\left(\bar{s}_{n,j}\geq s_{n,j}\right) - \left[\widehat d_{a_n,j}\right]^+ \right]^+ \label{eq:reliability_aware_objective}
\\ 
\mathrm{s.t.}\quad & \forall n:\; a_n\in\{1,\ldots,K\},
\\ & 
\forall n\in\mathbb{S}:\; \mathbf{c}_{a_n}\geq\mathbf{s}_n,
\label{eq_capabilities_filter}
\\ & 
\{\bar{\mathbf{s}}_n\}_{n\in\mathbb{S}} = \Gamma\!\left(\{a_n\}_{n\in\mathbb{S}}\right) 
\end{align} \end{subequations}
where \eqref{eq_capabilities_filter} defines the initial filtering of impossible allocations based on service capabilities, which can reduce the workload on the DT. 
The outer positive-part operator keeps the corrected fulfillment measure nonnegative. The component $\left[\widehat d_{a_n,j}\right]^+$ makes the systematically optimistic service DTs less likely to be selected in subsequent allocation decisions.


Recall that, in general, for $K$ services and $|\mathbb{S}|$ there are $K^{|\mathbb{S}|}$ allocations to be tested by the DT.
However, there are methods besides capabilities that can reduce the optimization space. For example, if a given service cannot support two intents, adding another intent will certainly not allow the QoS requirements of the intents to be met. Moreover, some services can block each other. For instance, an MNO can provide a 6G network using either 100 MHz or 200 MHz of bandwidth, but cannot use both simultaneously. Such heuristics can reduce the DT workload. In addition, some reinforcement learning tools, such as policy factorization, can further decrease optimization complexity, though this comes at the cost of reduced intent-service assignment quality. While this paper presents the optimal performance of the proposed framework, the analysis of suboptimal solutions is left for future research. 

Most importantly, note that the above objective function inherently favors existing allocations. The current intent-service allocation, if it meets the KPI requirements, should have a smaller variance and therefore a higher probability of exceeding the KPI threshold $s_{n,j}$ than a new intent-service pair. Note that the above objective function can be further refined by explicitly penalizing intent-to-service reassignment. Other objectives are also possible, such as accounting for the energy efficiency of the provided services. In this paper, the main goal is to maximize the probability of service provisioning at the required QoS level.

Finally, observe that although the above optimization problem always selects a single solution, it is possible that the probability of meeting requirements for the "worst-case" intent, i.e., $\min_{n,j} \Pr({\bar{s}} _{n,j} \geq s_{n,j})$, is too low. In such a case, IDMO should inform the intent provider that it is unable to support its request. Then the assignment problem can be executed again without the rejected intent.  Optionally, the intent provider can modify its intent by reducing its requirements, a process commonly known as "negotiations" \cite{tmforum_tr290_intent_common_model}.

\section{Experimental scenario and simulation results}
\label{sec:scenario_results}

We consider a $3\times4$ km area served by two MNO services with the parameters listed in Table~\ref{tab:services}. Service 1 prioritizes capacity, whereas Service 2 prioritizes coverage. The advertised capabilities follow the 3GPP TS 28.541 ServiceProfile attributes Table~\ref{tab:services}.
Hence, the IDMO uses them only for coarse filtering of intent-service pairs.
The maximum bitrate is obtained as the sum of the per-cell peak data rates computed using the approximate formula given in 3GPP TS~38.306~\cite{3gpp38306}, Sec.~4.1.2. For both services, we assume numerology $\mu=0$ (15 kHz subcarrier spacing), a single MIMO layer ($v=1$), 64QAM ($Q_m=6$), the maximum code rate $R_{\max}=948/1024$, a scaling factor $f=1$, and a downlink overhead $OH=0.14$. The number of PRBs follows from the channel bandwidth, i.e., $N_{\mathrm{PRB}}=106$ for 20~MHz (Service~1) and $N_{\mathrm{PRB}}=79$ for 15~MHz (Service~2). This results in approximately 85.1~Mbps and 63.4~Mbps per cell, which gives 766~Mbps for Service~1 (9 cells) and 190~Mbps for Service~2 (3 cells), as listed in Table~\ref{tab:services}. The maximum number of users reflects the admission limit configured by the MNO. In our case, it equals the maximum bitrate divided by a nominal throughput of 100 kbps per user. As a next step, the IDMO invokes the DT to provide high-granularity metrics.

\begin{table}[htp!]
    \centering
    \caption{Services Parameters}
    \label{tab:services}
    \begin{tabular}{lcc}
        \toprule
        Parameter & Service 1 & Service 2 \\
        \midrule
        Number of cells before/after earthquake & 9/5 & 3/3 \\
        Carrier frequency [MHz] & 2100 & 700 \\
        Bandwidth [MHz] & 20 & 15 \\
        TX power [dBm] & 40 & 30 \\
        Antenna height [m] & 25 & 25 \\
        \makecell[l]{ Capability: Max. users\\before/after earthquake}
         & 7660/4260 & 1900/1900 \\
        Capability: Min. Outage [\%] & 0 & 0 \\
        \makecell[l]{Capability: Max. Bitrate\\before/after earthquake [Mbps]} & 766/426 & 190/190 \\
        \bottomrule
    \end{tabular}
\end{table}

In this scenario, two intents are registered with IDMO: Intent~1 at 0~s and Intent~2 at 1000~s. Both are received as natural language and translated by the Intent Handler into a RAN-specific RFS. This RFS is simplified and includes only four parameters: the number of UEs $N$, the minimum total bit rate $R_{\min}$, the maximum allowed outage $O$, and the UE speed $v$ (Table~\ref{tab:intents}). To bridge the gap between human language and network KPIs, a prompt engineering approach, as described in \cite{dinhtowards2025}, is used, specifically Zero-shot Prompting combined with Role Prompting. The translation is performed by the Qwen3.5-397B-A17B model \cite{qwen35}. The model is queried with a temperature of 0, and its output is constrained by a JSON schema (guided decoding), so every response is a syntactically valid object with a fixed set of fields.

\begin{table}[htp!]
    \centering
    \caption{Intents Parameters (RFS)}
    \label{tab:intents}
    \begin{tabular}{lcc}
        \toprule
        Parameter & Intent 1 & Intent 2 \\
        \midrule
        Number of UEs & 50 & 45 \\
        Minimum total bitrate[Mbps] & 95 & 85.5 \\
        UE speed [$km/h$] & 5.4 & 5.4 \\
        Maximum Outage [\%] & 5 & 5 \\
        \bottomrule
    \end{tabular}
\end{table}

The role of the LLM is deliberately limited to understanding natural language. It extracts explicitly stated values (number of users, bitrate, speed, availability, or outage) and classifies the request into a service profile and a mobility class. It does not compute any KPIs itself. All numerical mapping is performed deterministically by the Intent Handler using the inference rules shown in Listing~\ref{lst:full_prompt}. If no bitrate is specified, the per-UE bitrate $r$ is taken from the service profile (4K video: 25~Mbps, HD video: 5~Mbps, gaming: 10~Mbps, web/office: 2~Mbps). Human users without a specific service mentioned, such as tourists and visitors in Intents~1 and~2, are assigned the generic web/office profile (2 Mbps), which corresponds to browsing and mobile app traffic (e.g., guide or navigation apps). Similarly, the speed $v$ follows from the mobility class (stationary: 0 km/h, pedestrian: 5.4 km/h, vehicular: 50 km/h), unless stated explicitly, and the outage follows from the required availability $A$ as $O = 100\% - A$. Instead of passing the per-UE bitrate directly, the minimum total required bitrate is calculated from the number of UEs, the per-UE bitrate, and the maximum outage:
\begin{equation}
    R_{\min} = N \cdot r \cdot (1 - O/100).
    \label{eq:rmin}
\end{equation}
For Intent~1, this gives $50 \cdot 2 \cdot 0.95 = 95$~Mbps, and for Intent~2, $45 \cdot 2 \cdot 0.95 = 85.5$~Mbps (Table~\ref{tab:intents}).


\noindent
\begin{lstlisting}[float,
    caption={System prompt, the two user intents and the LLM output resulting in Intent RFSs from Table~\ref{tab:intents}},
    label={lst:full_prompt},
    basicstyle=\footnotesize\ttfamily,
    breaklines=true,
    breakindent=0pt,
    frame=single,
    framerule=0.4pt,
    xleftmargin=3pt,
    xrightmargin=3pt,
    resetmargins=true,
    columns=fullflexible,
    keepspaces=true,
    showstringspaces=false
]
You are a 5G/6G network intent extractor. Read a service request written in natural language (any language) and fill the JSON schema. Do NOT compute derived values - only report what is stated, and classify where asked.

FIELDS
- user_count: number of users/UEs/devices; sum groups if several are given. null if not stated.
- service_type: video_4k | video_hd (video calls, webinars) | gaming | web_office (web browsing, office work, messaging, maps, mobile apps e.g. tourist or museum guide apps; also the default for people - tourists, visitors, employees - when no specific service is mentioned) | unknown (machine-type devices with unspecified traffic)
- mbps_per_user / total_mbps: ONLY if a number is given; convert to Mbps.
- mobility: stationary (seated, at home, fixed positions) | pedestrian (walking, on foot, touring, moving around a venue) | vehicular (cars, buses, trains) | unknown
- ue_speed_kmh: ONLY if stated explicitly (convert m/s to km/h).
- availability_percent / outage_percent: if stated.
Use null for anything not stated. Never guess numbers that are not in the text.
-------------------------------------------
USER INTENT 1:
"We need to cover 50 tourists visiting the old town museum on foot. We aim for 95% availability."
LLM OUTPUT:
{"user_count": 50, "service_type": "web_office", "mbps_per_user": null, "total_mbps": null, "mobility": "pedestrian", "ue_speed_kmh": null, "availability_percent": 95, "outage_percent": null}
-------------------------------------------
USER INTENT 2:
"A group of 45 visitors will walk around the botanical garden using a mobile guide app. An outage of up to 5% is acceptable."
LLM OUTPUT:
{"user_count": 45, "service_type": "web_office", "mbps_per_user": null, "total_mbps": null, "mobility": "pedestrian", "ue_speed_kmh": null, "availability_percent": null, "outage_percent": 5}
\end{lstlisting}

To quantify the accuracy of the translation step, the Intent Handler was evaluated on a labeled set of 28 intents. Fifteen of these are basic requests covering all service profiles and mobility classes, both explicit and inferred bitrates, and include a request in Polish. The remaining 13 are deliberately harder: numbers written as words (``a dozen'', ``forty-five''), availability expressed indirectly (``five nines'', ``at most one failure in twenty'', ``36 hours of downtime over 30 days''), distractor numbers (``the company has 300 employees, but only 40 will join''), unit conversions (kbps, Gbps, MB/s, m/s), negations, and machine-type devices with no known bitrate. Each intent was processed three times. A parameter was counted as correct if it was within 1\% of the ground truth, and a missing value (null) was counted as correct only when no value was expected. All 84 responses were valid JSON. The accuracy of each of the four RFS parameters and the exact-match rate (all four parameters correct) were both 100\%, and the outputs were identical across repetitions. These results indicate that, for the test set considered and the four core parameters, a refined zero-shot strategy with constrained output is sufficiently accurate to support the subsequent DT-based evaluation. The current schema assumes a single service profile per intent. Intents that combine user groups with different services (e.g., some users streaming 4K video while others browsing the web), as well as adversarial prompts, are outside the scope of this work. Handling these cases would require extending the schema to support a list of user groups or using more advanced techniques from the literature, such as few-shot learning.


We will show the strengths of the proposed orchestration by considering multiple (in-time) intent-service assignment processes, including intent-support recovery after a natural disaster. The considered scenario evolves with events:
\begin{itemize}
    \item \textit{t = 0s} - Registering Intent 1 and launching the DT-based intent-service assignment;
    \item \textit{t = 1000s} - Registering Intent 2 and launching the DT-based intent-service assignment;
    \item \textit{t = 2000s} - Earthquake, resulting in the failure of four Service 1 cells;
    \item \textit{t = 3000s} - Service recovery by launching the DT-based intent-service reassignment for improving the QoS.
\end{itemize}  

During each intent-service assignment process, it is first checked whether any service can reject any intent based on the capabilities vector as defined by (\ref{eq_capabilities_filter}). In our scenario, every intent can be served by any service, as shown by comparing the values in Table \ref{tab:intents} and Table \ref{tab:services}. Next, the performance of the services is evaluated in the DT. Recall that, due to ambiguity in intent mapping to the DT configuration, multiple DT instances (1,000 in our case) will run in parallel, each with a different set of random parameters to return KPI distributions. The observed KPIs in our case are \emph{potential bitrate} and outage values.
If a DT allocates only the bitrate required by an intent, services with substantially different available capacities may appear equivalent because both provide a bitrate close to the requirement. We therefore define the potential bitrate as the bitrate achievable by an intent if all resources unused by other intents are allocated to it, assuming that the other intents receive the resources required by their $\mathbf{s}_n$
vectors. While UEs in the outage are counted by the outage metric, the potential bitrate sums only the bitrate of users not being in the outage state. This metric reveals the available performance margin and enables services with different capacities to be distinguished. A larger margin implies a higher probability of satisfying the requirement under uncertain network conditions. The second analyzed KPI is outage, defined as the fraction of UEs that achieve less than 95\% of their target bitrate. It may result from insufficient resources or lack of coverage. The number and speed of UEs determine the DT configuration and are therefore not treated as DT outputs.

\subsection{Simulator specification}
The proposed framework is evaluated using the time-driven system-level simulator introduced in \cite{kryszkiewicz2026largescalewirelessnetworkmanagement}, with a simulation step of 1~s. At each step, UEs move according to a random walk model, path loss, shadowing, and SINR are updated, and each UE connects to the allowed BS providing the highest RSRP. The BSs allocate resource blocks using Round-Robin scheduling until the UE throughput requirements are met, and the achieved bitrates are collected as KPIs. The simulation parameters are summarized in Table~\ref{tab:simulation_parameters}. The 12 BSs form a $3\times4$~km grid with an inter-site distance of 1000~m, while uniformly distributed UEs move at 5.4~km/h in random directions. Path loss follows the 3GPP Urban Micro model~\cite{3gpp_tr38901} with spatially correlated shadowing. Inter-cell interference is calculated explicitly, and the SINR is mapped to bitrate using the truncated Shannon model~\cite{3gpp_tr38803}. The simulator represents both the physical network and the DT, but the DT has incomplete knowledge of the physical network, as discussed in Section~\ref{sec_DT_output}. Since both use the same underlying simulator, $\widehat d_{k,j}^{(\ell)}$ should converge to zero. Nonzero systematic errors are studied in Sections~\ref{sec_tuning_estimator} and~\ref{sec_with_systematic_error}. Elsewhere, $\widehat d_{k,j}^{(\ell)}=0$ is assumed to isolate other effects. The main source of uncertainty is the UE location. The physical network knows the UE locations and corresponding path losses, whereas the DT distributes UEs randomly unless information from an active or recently used intent-service allocation is available. Each candidate configuration is evaluated using 1000 random DT realizations. Each realization lasts 20~s, with the last 10 measurements, i.e., after stabilization, used for decisions, yielding 10000 samples per KPI distribution.

\begin{table}[htp!]
    \centering
    \caption{Simulation Parameters}
    \label{tab:simulation_parameters}
    \begin{tabular}{lc}
        \toprule
        Parameter & Value \\
        \midrule
        Number of base stations & 9 (Service 1) + 3 (Service 2) \\
        Grid dimensions [km] & $3 \times 4$ \\
        Inter-site distance [m] & 1000 \\
        Antenna height [m] & 25 \\
        \midrule
        Carrier frequency [MHz] & 2100 (Service 1), 700 (Service 2) \\
        Bandwidth [MHz] & 20 (Service 1), 15 (Service 2) \\
        TX power [dBm] & 40 (Service 1), 30 (Service 2) \\
        \midrule
        Number of UEs [Intents] & 50 (Intent 1) + 45 (Intent 2) \\
        Mobility model & Random walk \\
        UE speed [km/h] & 5.4 \\
        \midrule
        Propagation model & Urban Micro \\
        Shadowing model & 2D correlated fading \\
        Bitrate model & 3GPP TR 38.803 \\
        Allocation Model & Round Robin \\
        \bottomrule
    \end{tabular}
\end{table}



\subsection{Probability prediction error estimator tuning}
\label{sec_tuning_estimator}

\begin{figure}[htp!]
\centering
\includegraphics[trim={0cm 0cm 0cm 0cm}, clip,width=0.489\textwidth]{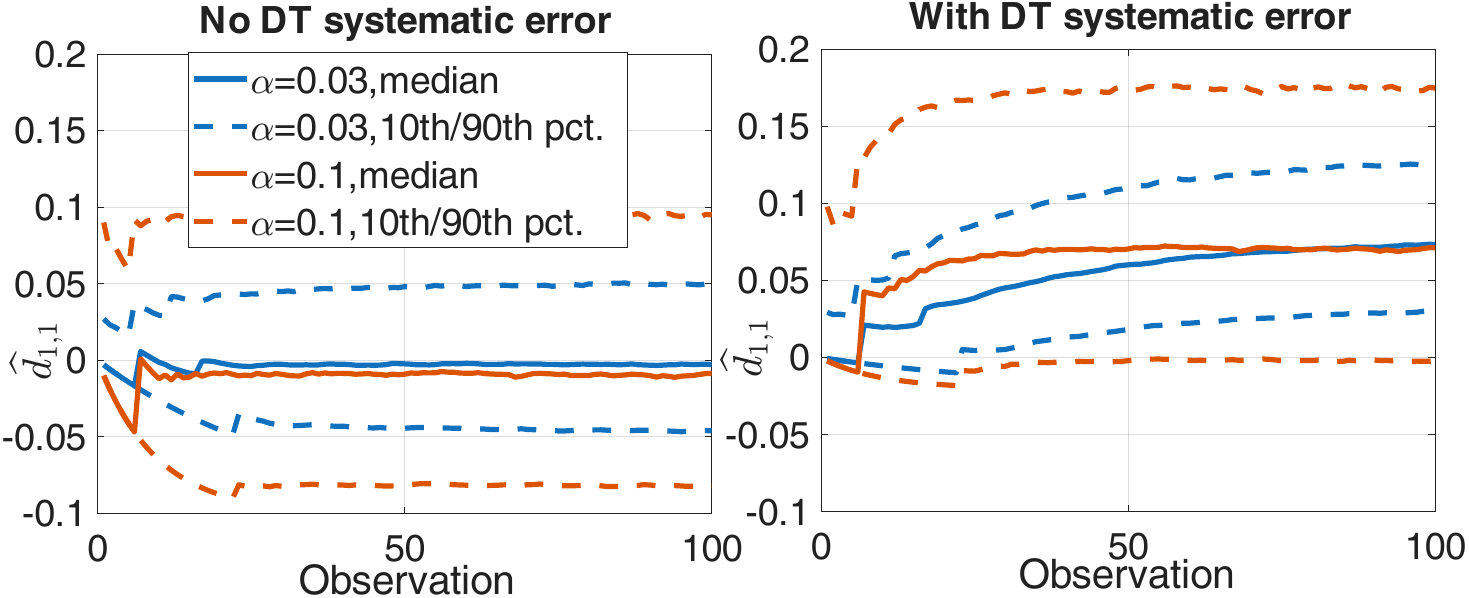}
\caption{Probability prediction error $\widehat d_{1,1}^{(\ell)}$ for DT with and without systematic error when $\alpha \in \{0.1, 0.03\}$ with median, 10th/90th percentiles obtained over 1000 parallel runs.}
\label{fig:probability_prediction_error}
\end{figure}

As discussed in Sec.~\ref{sec_problem_formulation}, the estimate $\widehat d_{k,j}^{(\ell)}$ allows the Orchestrator to compensate for systematic errors in a service DT, caused, for example, by model simplifications or deliberate optimism. The update factor $\alpha$ determines the trade-off between rapid adaptation and the suppression of fluctuations arising from randomness in both the environment and the intent. Two scenarios are considered. In the first case, both the DT-predicted and physical-network probabilities of satisfying the KPI are 0.9. In the second case, these probabilities are 0.98 and 0.9, respectively, corresponding to a systematic DT error of 0.08. For each scenario, 1,000 independent estimator runs are performed, with $y_{1,1}$ generated according to the physical network fulfillment probability. The estimator is initialized with $\widehat d_{1,1}^{(0)} = 0$. Fig.~\ref{fig:probability_prediction_error} presents the median, as well as the 10th and 90th percentiles, for $\alpha = 0.1$ and $\alpha = 0.03$. Without systematic DT error, the median estimate remains close to zero. However, the 90th percentile reaches approximately $0.05$ for $\alpha=0.03$ and $0.1$ for $\alpha=0.1$. Because positive deviations may temporarily reduce the probability of selecting a correctly calibrated service in \eqref{eq:reliability_aware_objective}, choosing $\alpha = 0.03$ is preferable due to its lower variability. When the DT overestimates the fulfillment probability by $0.08$, the median estimate reaches this value after approximately 60 observations for $\alpha = 0.03$. For $\alpha = 0.1$, this happens several times faster, illustrating the trade-off between estimator stability and adaptation speed. Since relatively small systematic errors are expected, $\alpha = 0.03$ is used in Sec. \ref{sec_with_systematic_error}.


\subsection{DT variance validation}

\begin{figure}[htp!]
\centering
\includegraphics[trim={0cm 0cm 0cm 0cm}, clip,width=0.49\textwidth]{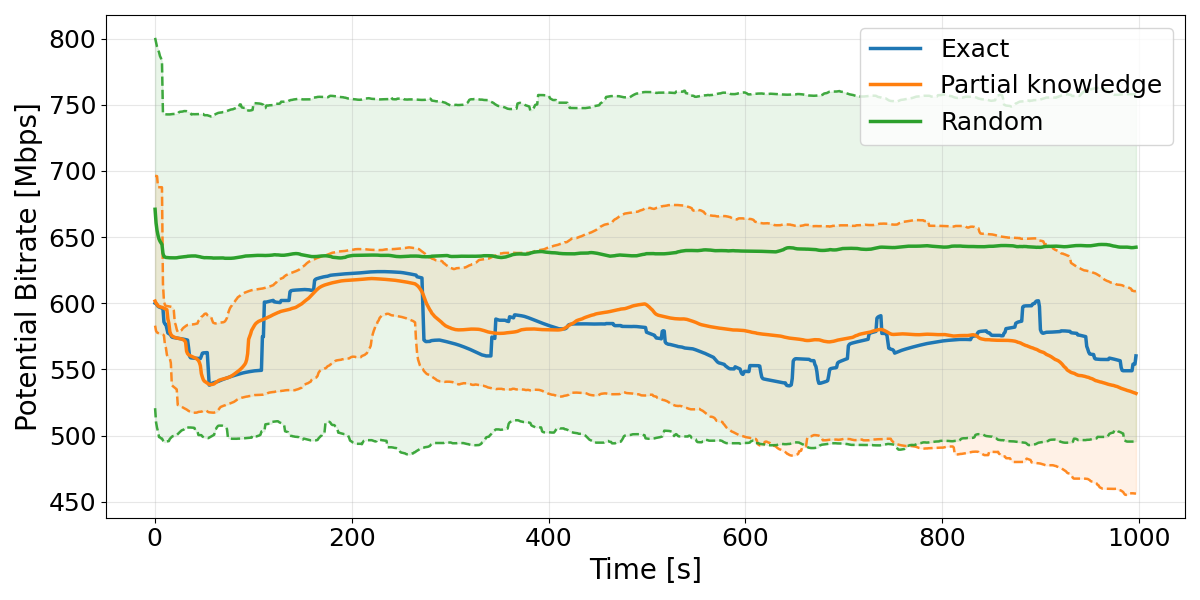}
\caption{Time-varying Potential Bitrate across the real network ("Exact") and 2 DT realizations (with different initial knowledge). Median, 1st/99th percentiles shown. 
}
\label{fig:variance_10_90}
\end{figure}

First, to justify our DT inaccuracy modeling discussed in Section \ref{sec_DT_output} as the basis for our intent-service mapping, we present the potential bitrate as the DT state diverges from the physical-network state. In the following analysis, we assume that Intent 1 is allocated to Service 1. The results are shown in Fig.~\ref{fig:variance_10_90}, where the median potential bitrate, along with the 1st and 99th percentiles, is plotted over time for the real network (“Exact”) and two DT configurations: “Partial Knowledge” and “Random" scenarios. In the Partial Knowledge setup, the DT is initialized with real-world UE positions at time 0 s. Subsequently, UE mobility follows a random process that differs from the Exact scenario. In contrast, the Random case assumes no prior knowledge, with UEs randomly distributed from the start.

The Random scenario shows a relatively stable median and 1st/99th percentiles over time, with a difference of about 125 Mbps between the median and the 1st percentile. In contrast, the Partial Knowledge scenario shows increasing variability over time, indicating that uncertainty grows as the system evolves away from the initially known state due to divergence between actual UE mobility and its DT-based approximation. Although even after 100 s the difference between the 1st and 99th percentiles is still around 90 Mbps, it is much smaller than in the “Random” case. This highlights that incorporating partial knowledge enhances the DT’s ability to represent the real-world scenario, particularly shortly after a given intent-service is used, potentially leading to more reliable decision-making by the IDMO.


\subsection{IDMO allocation results}

\begin{figure}[htp!]
\centering
\includegraphics[trim={0cm 0cm 0cm 0cm}, clip,width=0.49\textwidth]{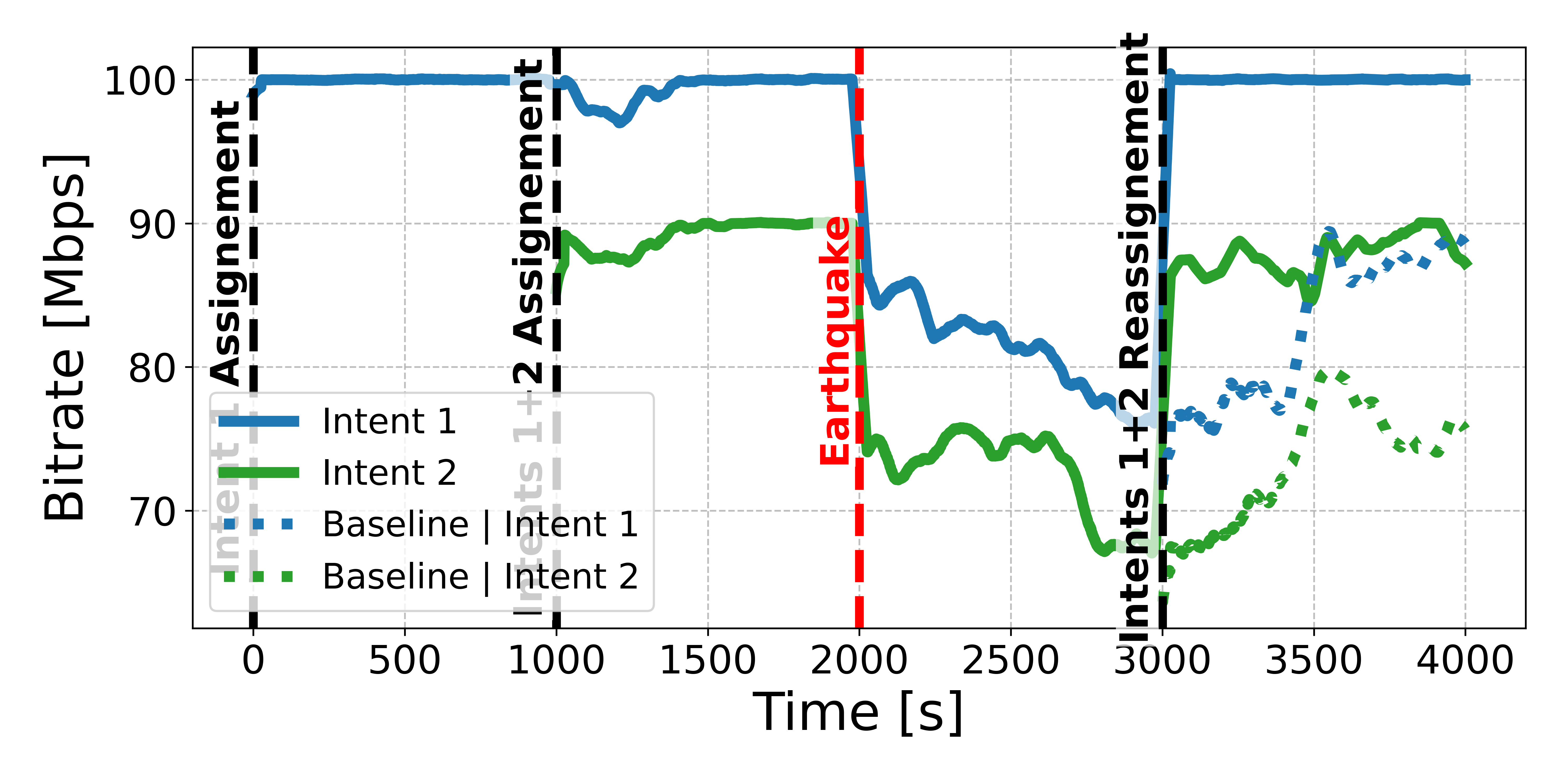}
\caption{Bit rates of intents over time from real network.}
\label{fig:main_simulation_bitrate}
\end{figure}
\begin{figure}[htp!]
\centering
\includegraphics[trim={0cm 0cm 0cm 0cm}, clip,width=0.49\textwidth]{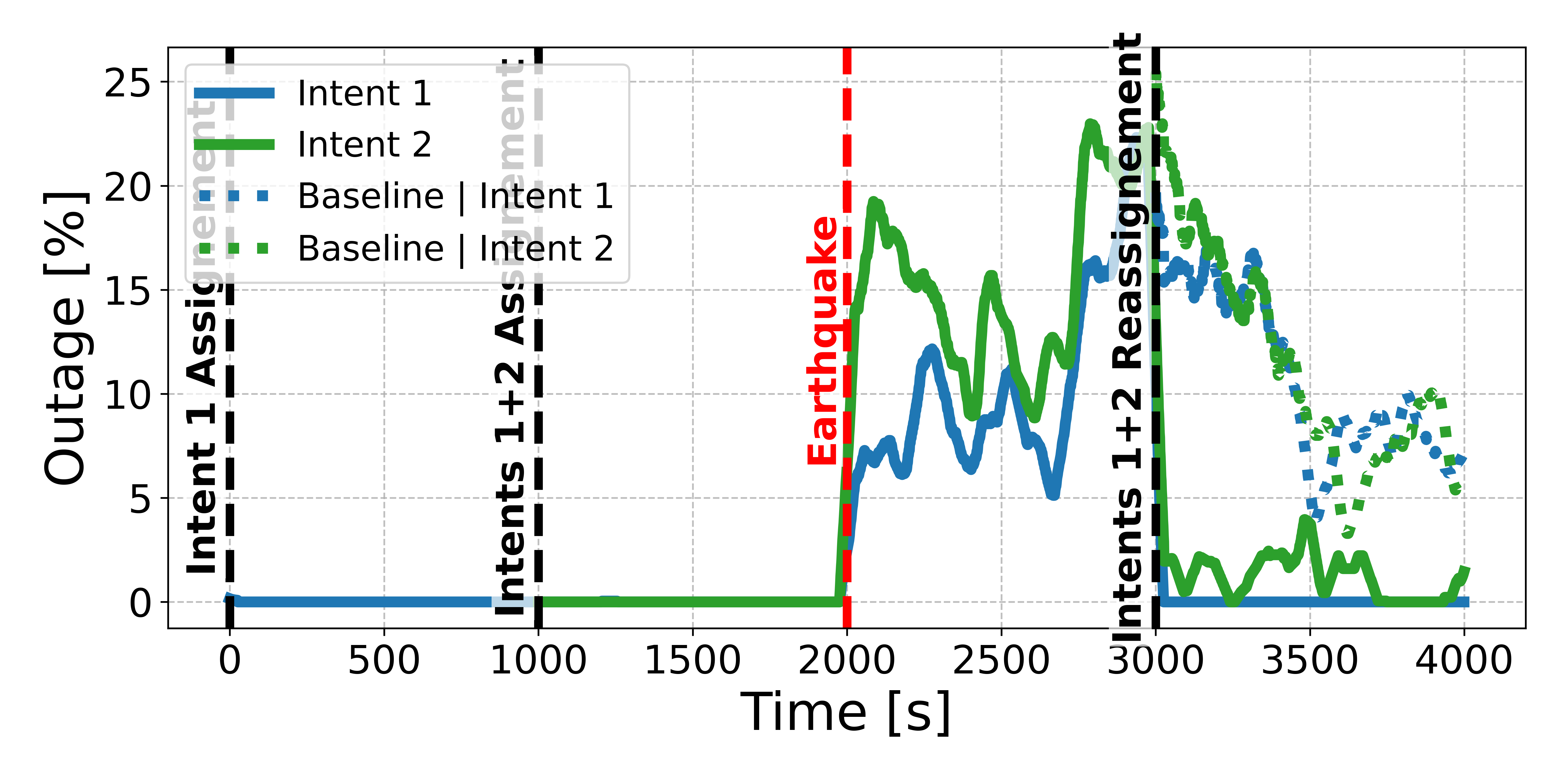}
\caption{Intents outage in time from the real network.}
\label{fig:main_simulation_outage}
\end{figure}

Now we analyze the results of the DT-based IDMO allocation. We compare it against the \emph{Baseline} scheme, which allocates intents to the first service that meets their requirements according to the capabilities defined in Table \ref{tab:services}. Note that we assume the Service 1 provider can rapidly update its capabilities after an earthquake. First, the actual network KPIs for the considered scenario are shown in Fig.~\ref{fig:main_simulation_bitrate} and Fig.~\ref{fig:main_simulation_outage} for bitrate and outage, respectively. After this analysis, we will focus on each IDMO assignment problem and present the DT outputs. The vertical lines represent global events affecting the system. At $t = 0$ s, intent 1 is registered in IDMO. IDMO then launches the assignment process: for each possible intent-service allocation, a DT runs (composed of 1000 random instances to obtain KPI distributions). Based on the DT output, IDMO selects the intent-service allocation using \eqref{eq_opt_problem}. Initially, assigning intent 1 to Service 1 is determined to be optimal. Similarly, at $t = 1000$ s, intent 2 is introduced and the DT is executed again, allocating both intents to Service 1, i.e., $\mathbf{a} = [1, 1]$. From $t = 0$ s to $t = 2000$ s, the outage for both intents remains 0, while the system bitrate is maintained at 100 Mbps for intent 1 and 90 Mbps for intent 2 during their respective activity periods. Therefore, all UE demands are satisfied during this period. At $t = 2000$ s, an earthquake scenario is simulated, causing four cells of Service 1 to become unavailable. Following the earthquake, the outage increases for both intents and exceeds $10\%$ during portions of the degraded-operation interval while their bitrates fall below the required levels, as shown in Fig. \ref{fig:main_simulation_bitrate} and Fig. \ref{fig:main_simulation_outage}. To illustrate the impact of the earthquake, the original intent-service assignment is retained from $t = 2000$ s to $t = 3000$ s. During this interval, both intents remain assigned to the damaged Service 1. At $t = 3000$ s, a DT-assisted reassignment is applied, retaining Intent 1 on Service 1 and moving Intent 2 to Service 2. The deliberately imposed interval illustrates degraded operation before recovery and does not represent a measured computation or service-recovery duration. Based on the IDMO decision to allocate Intent 1 to Service 1 and Intent 2 to Service 2, the outage for Intent 1 is reduced to $0\%$, and the outage for Intent 2 is also decreased below the required level, improving overall system performance. These results demonstrate that DT-supported IDMO can improve service quality after infrastructure failure by selecting a more suitable intent-service assignment. However, the experiment does not measure the time required for failure detection, DT evaluation, and reassignment execution. The Baseline solution allocates both intents to Service 1, overlapping with the DT-aided assignment until $t = 3000$ s. However, after $t = 3000$ s, both intents remain with Service 1 because they meet the Service 1 capability specifications, resulting in significantly worse throughput and outage performance than the DT-aided assignment.

\begin{figure}[htp!]
\begin{subfigure}{0.24\textwidth}
    \includegraphics[width=\linewidth]{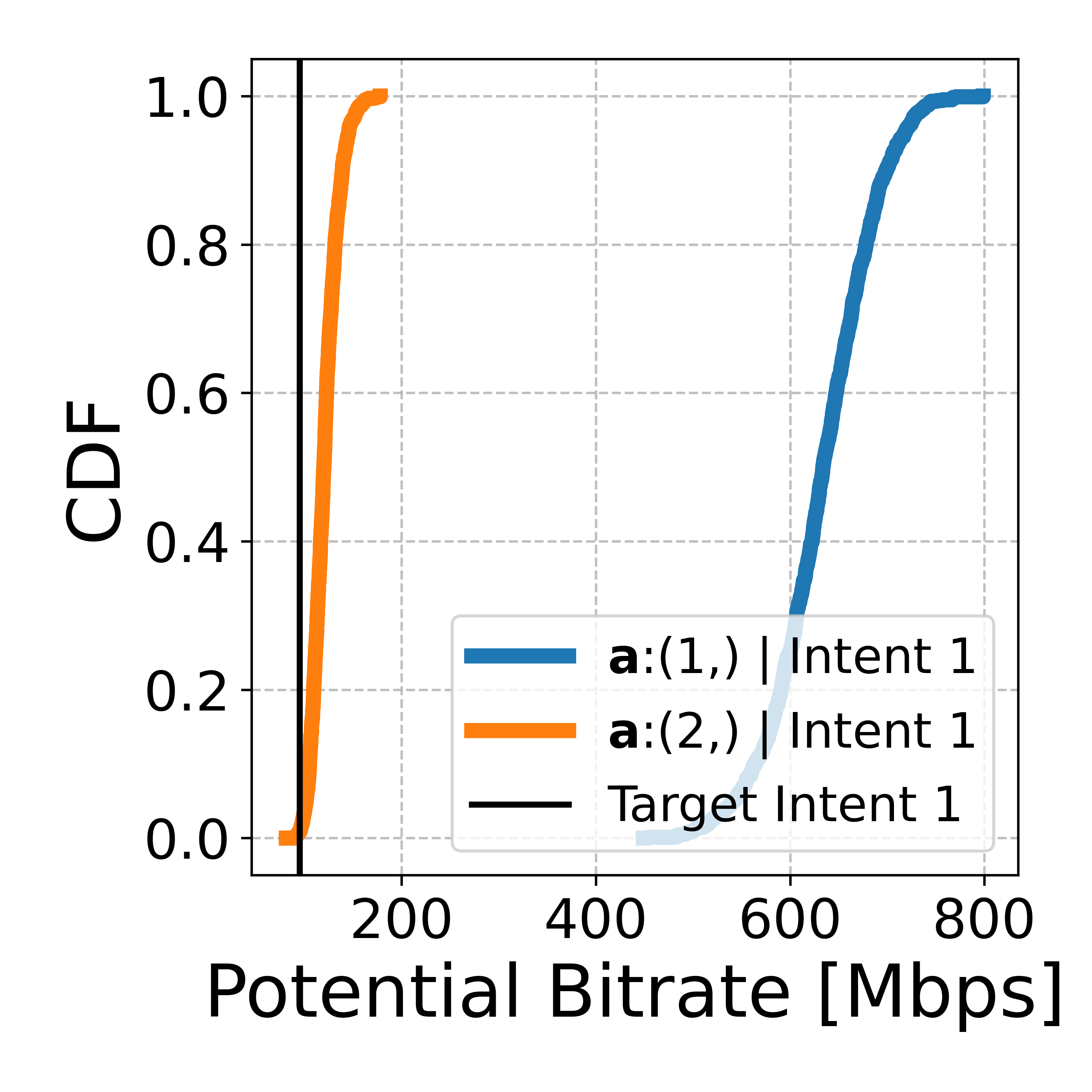}
    \caption{Bitrate}
    \label{fig:bitrate_0s}
\end{subfigure}
\begin{subfigure}{0.24\textwidth}
    \includegraphics[width=\linewidth]{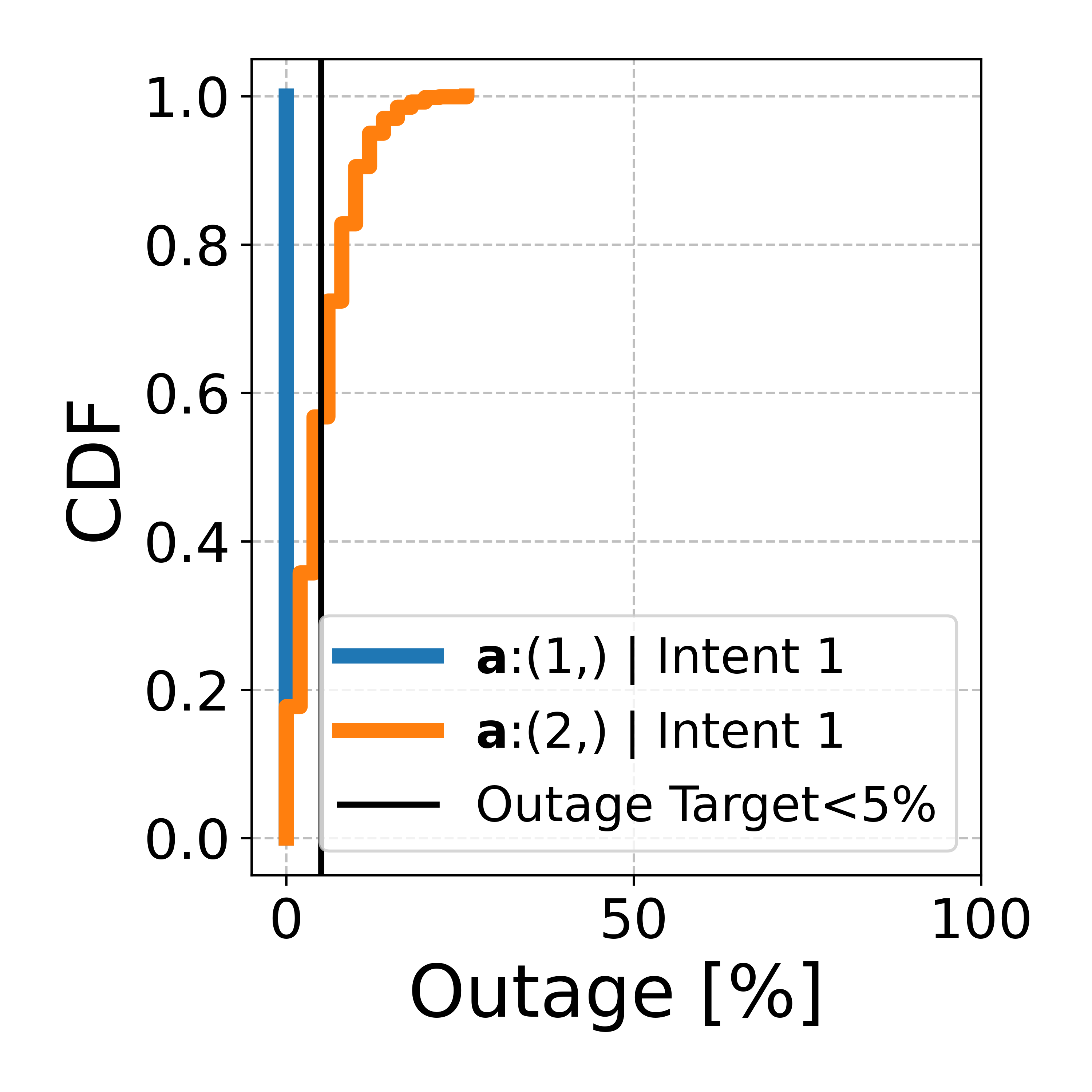}
    \caption{Outage}
    \label{fig:outage_0s}
\end{subfigure}
\caption{DT-estimated CDFs of potential bitrate and outage for the candidate allocations at $t=0$~s. Recall: $\mathbf{a}=[2]$ means Intent 1 is allocated to Service 2. }
\label{fig:bitrate_outage_0s}
\end{figure}

We next analyze the allocation decisions made by IDMO using \eqref{eq:reliability_aware_objective}.
Fig.~\ref{fig:bitrate_outage_0s} presents the DT-estimated potential-bitrate and outage distributions for the first allocation at $t=0$~s. Service~1 satisfies both requirements (marked as vertical black lines) with a probability of approximately $100\%$. For Service 2, the probabilities of meeting the bitrate and outage requirements are $96.8\%$ and $52.9\%$, respectively. Therefore, the minimum fulfillment probability is $100\%$ for $\mathbf{a} = [1]$, compared with $52.9\%$ for $\mathbf{a} = [2]$, and IDMO assigns Intent 1 to Service 1.


\begin{figure}[htp!]
\centering
\includegraphics[trim={0cm 0cm 0cm 0cm}, clip,width=0.49\textwidth]{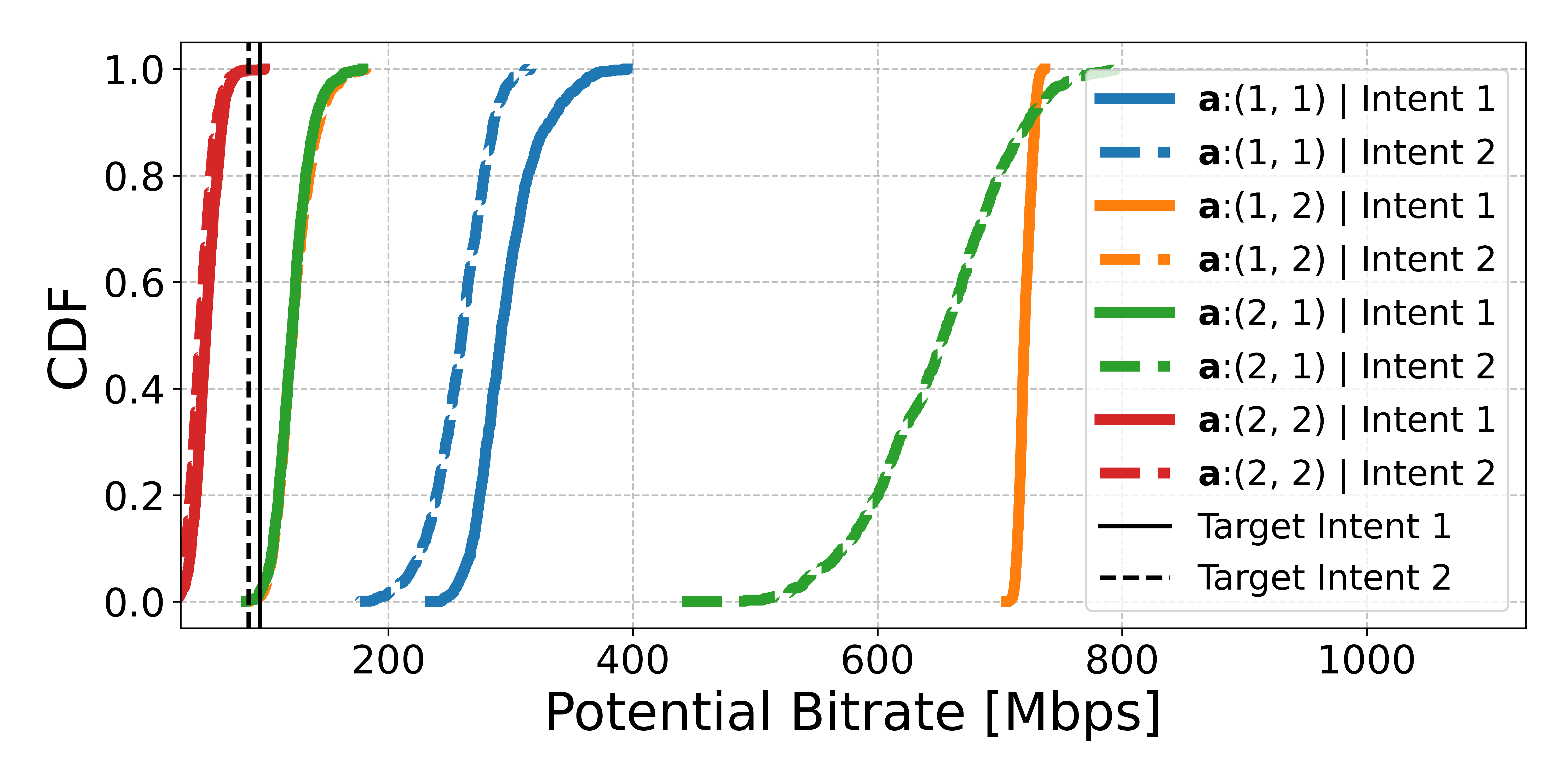}
\caption{DT-estimated CDFs of potential bitrate for the candidate allocations at $t=1000$~s.}
\label{fig:bitarate_100s}
\end{figure}

\begin{figure}[htp!]
\centering
\includegraphics[trim={0cm 0cm 0cm 0cm}, clip,width=0.49\textwidth]{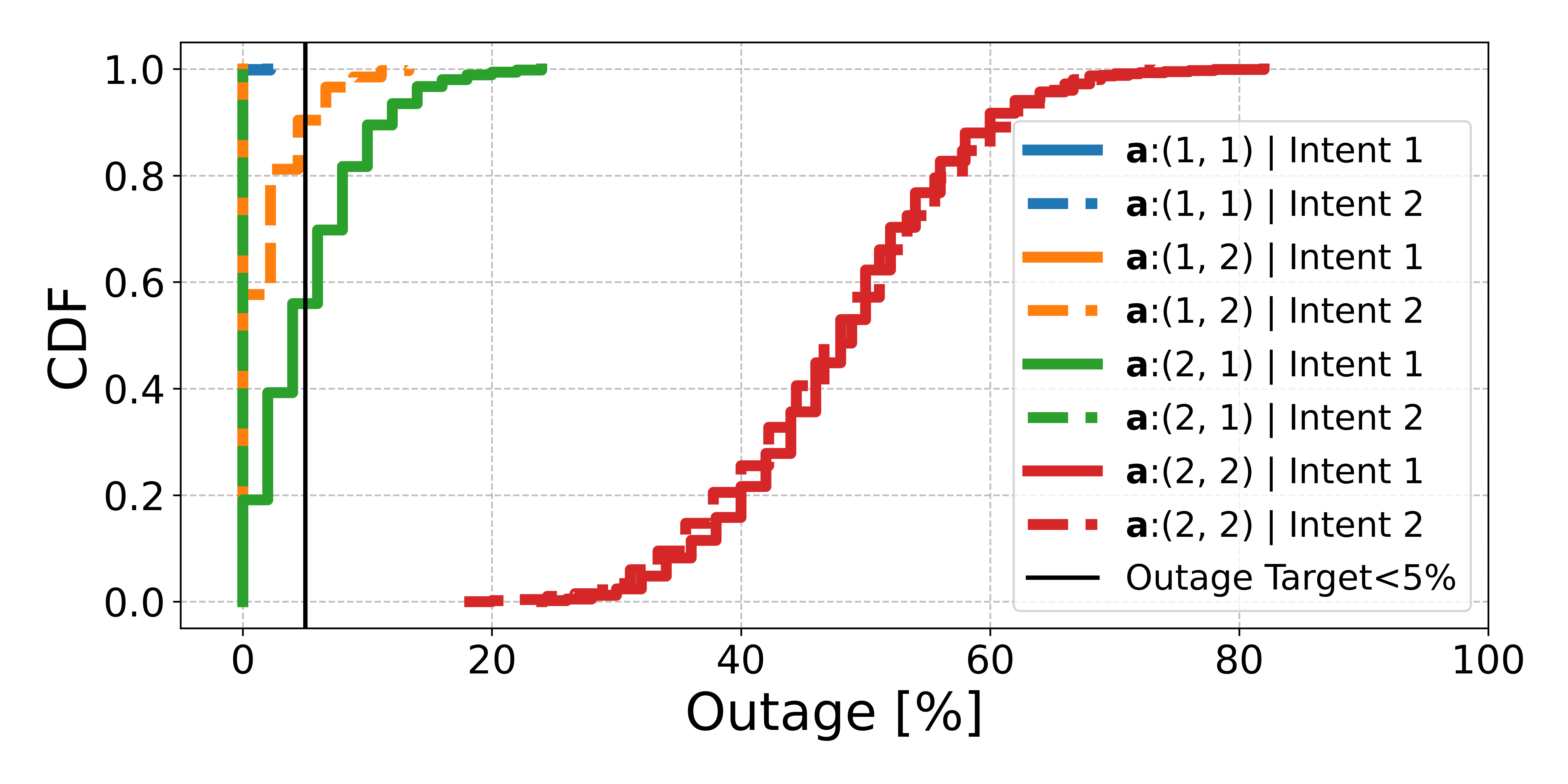}
\caption{DT-estimated CDFs of outage for the candidate allocations at $t=1000$~s.}
\label{fig:outage_100s}
\end{figure}

At $t=1000$~s, Intent~2 is introduced, and the DT evaluates four candidate allocations, as shown in Fig.~\ref{fig:bitarate_100s} and Fig.~\ref{fig:outage_100s}. Assigning both intents to Service~2, i.e., $\mathbf{a}=[2,2]$, yields only approximately $0.1\%$ and $0.1\%$ probabilities of satisfying the respective bitrate requirements. The remaining allocations satisfy the bitrate requirements with probabilities close to $100\%$. Outage is therefore the limiting KPI, and $\mathbf{a}=[1,1]$ provides the highest probability, close to $100\%$, that the outage is below 5\%. Both intents are assigned to Service~1.

\begin{figure}[htp!]
\centering
\includegraphics[trim={0cm 0cm 0cm 0cm}, clip,width=0.49\textwidth]{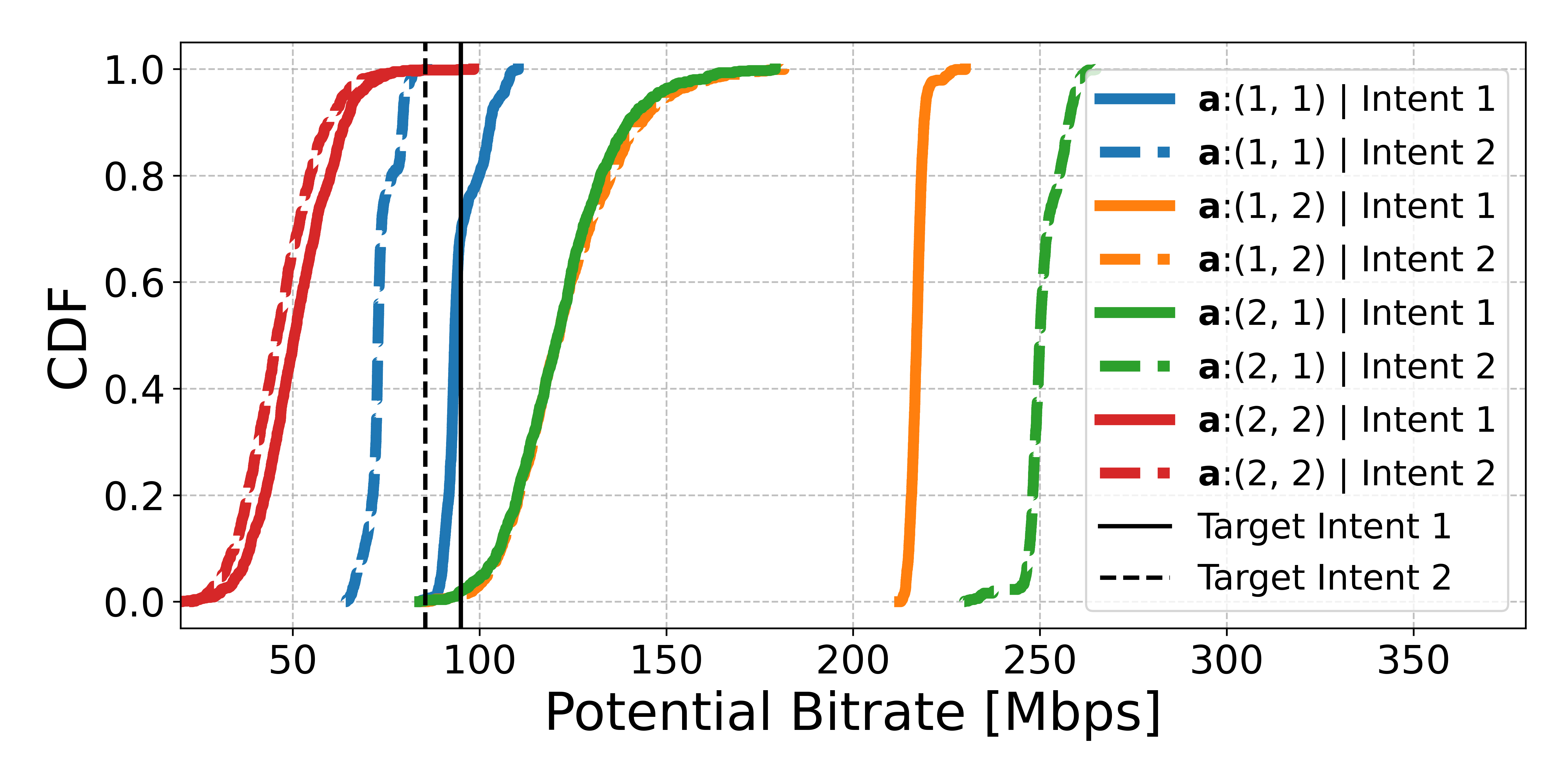}
\caption{DT-estimated CDFs of potential bitrate for the candidate allocations at $t=3000$~s.}
\label{fig:bitarate_3000s}
\end{figure}

\begin{figure}[htp!]
\centering
\includegraphics[trim={0cm 0cm 0cm 0cm}, clip,width=0.49\textwidth]{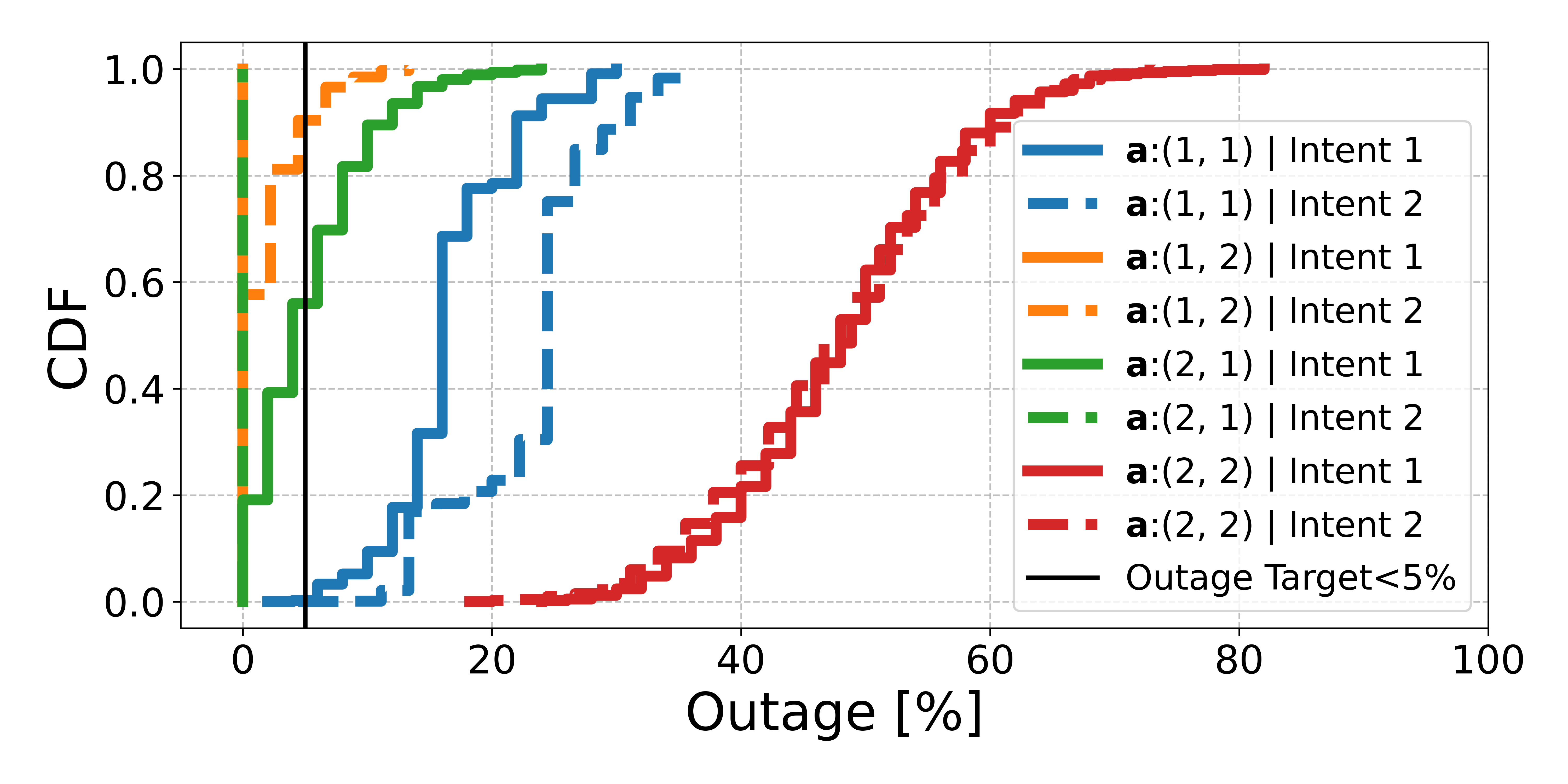}
\caption{DT-estimated CDFs of outage for the candidate allocations at $t=3000$~s.}
\label{fig:outage_300s}
\end{figure}

Following the earthquake at $t=2000$~s, the requirements of both intents are violated, as shown in Figs.~\ref{fig:main_simulation_bitrate} and~\ref{fig:main_simulation_outage}. Fig.~\ref{fig:bitarate_3000s} and Fig.\ref{fig:outage_300s} present the DT-estimated KPI distributions used to select the reassignment applied at $t=3000$ s. After the failure of four Service~1 cells, splitting the intents between the services provides the best performance. For $\mathbf{a}=[2,1]$, the limiting bitrate-fulfillment probability is $97.9\%$ for Intent~1, whereas for $\mathbf{a}=[1,2]$ it is $99.9\%$ for Intent~2. The outage results favor $\mathbf{a}=[1,2]$, for which the limiting probability of satisfying the outage requirement is $90.4\%$. IDMO therefore retains Intent~1 in Service~1 and reallocates Intent~2 to Service~2.  Additionally, the benefit of partial DT knowledge is evident when comparing Figs.~\ref{fig:bitrate_0s} and~\ref{fig:bitarate_100s}. For Intent~1 assigned to Service~1, the difference between the 99th and 1st percentiles decreases from $135.0$~Mbps at $t=0$~s, when no prior UE location information is available, to $14.5$~Mbps at $t=1000$~s, when the DT has information from the active allocation.

\subsection{ DT with systematic error}
\label{sec_with_systematic_error}

This section illustrates the assignment process when a systematic DT error is present. To this end, we define a simplified scenario with a single intent consisting of 90 UEs, each requiring 2 Mbps, and two candidate services: Service 1, deployed on one BS, and Service 2, deployed on four BSs. To simplify the presentation, only total throughput is used as a KPI. A systematic, potentially intentional error is introduced into Service~1 by multiplying its DT-estimated potential bitrate by 2. The DT-based assignment is performed every 10~s.

\begin{figure}[htp!]
\centering
\includegraphics[trim={0cm 0cm 0cm 0cm}, clip,width=0.49\textwidth]{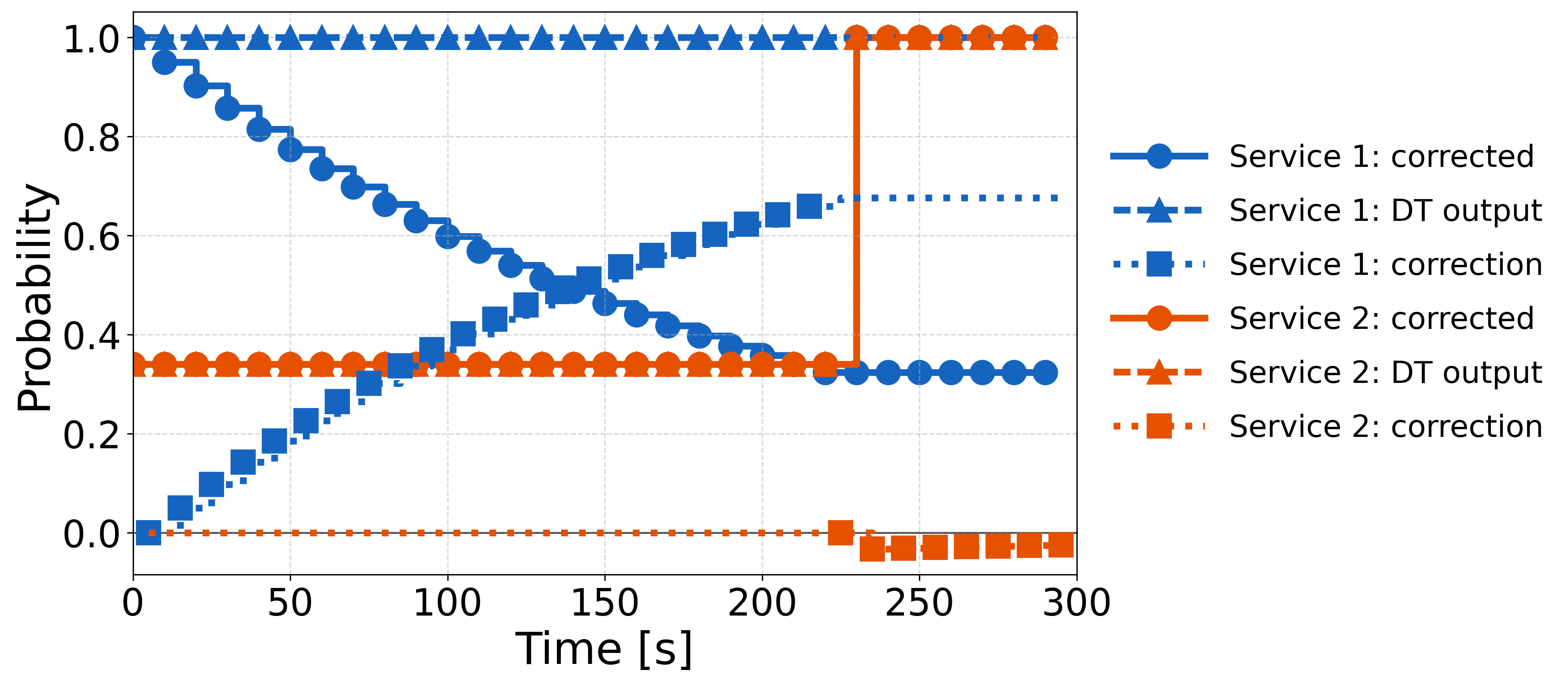}
\caption{Metrics used for intent-service assignment using
(\ref{eq:reliability_aware_objective}): DT output, correction (estimated DT
error), and corrected prediction, for assignments performed every 10~s.
Markers denote the moments at which new values are computed. Service~1 has an
injected DT error (potential bitrate increased by a factor of 2).}
\label{fig:DT-error}
\end{figure}

Fig.~\ref{fig:DT-error} shows, for both services, the uncorrected
probability of achieving the required bitrate, $\Pr\!\left(\bar{s}_{n,j}\geq s_{n,j}\right)$, denoted as \emph{DT output}; the estimated error $\widehat d_{a_n,j}$, denoted as \emph{correction}; and the corrected prediction $\Pr\!\left(\bar{s}_{n,j}\geq s_{n,j}\right) - \left[\widehat d_{a_n,j}\right]^+$, denoted as \emph{corrected}. Since the DT overestimates the performance of Service~1, its correction grows over time, and the corrected metric decreases accordingly. At approximately $t = 220$~s, the corrected metric for Service~1 falls to about 0.32, below the prediction of 0.34 for Service~2, and the framework switches the intent to Service~2. From the next assignment (at $t = 230$ s), the DT prediction for Service~2 increases to 1, as a result of the reduced intent-service ambiguity discussed in the previous subsection. After the switch, the correction for Service~1 remains constant at about 0.68, since no new observations of this service are collected. The correction of Service 2 becomes slightly negative after the switch. This is caused by $\Pr\!\left(\bar{s}_{n,j}\!\geq\! s_{n,j}\right)\!-\!y_{n,j}$ being around -0.5 after the first allocation decision update. The correction decays toward zero as subsequent predictions match the observations. Since only the positive part of the estimate is used in (\ref{eq:reliability_aware_objective}), this temporary negative value does not increase the allocation score of Service 2. This scenario shows that if the DT exhibits a systematic error, the proposed framework can detect and correct it.

\section{Conclusion}
\label{sec:conclusions}

This paper presented a DT-assisted framework for intent-service assignment in 6G service-based architectures. By integrating the DT into the orchestration loop, the framework enables more resource-efficient decisions while accounting for service interactions. Explicitly modeling DT prediction uncertainty removes the assumption of a perfect DT and improves the framework's applicability to real deployments. Simulations showed that DT randomization and probabilistic decision-making support reliable intent-service orchestration and recovery after network failures. Future work will explore scalable suboptimal solutions, such as reinforcement learning-based assignment, for large service and intent catalogs. 


%


\ifCLASSOPTIONcaptionsoff
  \newpage
\fi



\bibliographystyle{IEEEtran}
\bibliography{biblio}
\end{document}